\documentclass[pdflatex,sn-mathphys-num]{sn-jnl}

\usepackage{graphicx}%
\usepackage{multirow}%
\usepackage{amsmath,amssymb,amsfonts}%
\usepackage{amsthm}%
\usepackage{mathrsfs}%
\usepackage[title]{appendix}%
\usepackage{xcolor}%
\usepackage{textcomp}%
\usepackage{manyfoot}%
\usepackage{booktabs}%
\usepackage{array}
\usepackage{algorithm}%
\usepackage{algorithmicx}%
\usepackage{algpseudocode}%
\usepackage{listings}%
\usepackage{subfig}%
\usepackage{tabularx}%

\usepackage{booktabs}%
\usepackage{makecell}%
\usepackage{adjustbox}%
\usepackage{pdflscape}%
\usepackage{float}%

\theoremstyle{thmstyleone}%
\theoremstyle{thmstyletwo}%

\theoremstyle{thmstylethree}%

\begin{document}

\providecommand{\fevone}{FEV\textsubscript{1}}      
\providecommand{\fevonepp}{FEV\textsubscript{1}PP}  
\providecommand{\expect}{\operatorname{E}}          
\providecommand{\var}{\operatorname{Var}}           
\providecommand{\normaldistn}{\operatorname{Normal}}
\providecommand{\unifdistn}{\operatorname{Uniform}}
\providecommand{\halfnormaldistn}{\operatorname{half\mbox{-}Normal}}
\providecommand{\histdata}{\mathcal{D}}
\providecommand{\assurance}{\mathrm{A}}
\providecommand{\differential}{\mathrm{d}}

\title[Article Title]{Trial design considerations in the face of overdispersion and time-varying event rates: a meta-regression approach with an application in asthma}


\author*[1]{\fnm{Bohua} \sur{Chen}}
\email{bohua.chen@med.uni-goettingen.de}

\author[1]{\fnm{Christian} \sur{R\"over}}
\email{christian.roever@med.uni-goettingen.de}

\author[2,3]{\fnm{Stefan} \sur{Andreas}}
\email{stefan.andreas@med.uni-goettingen.de}

\author[4]{\fnm{Christian} \sur{Taube}}
\email{christian.taube@uk-essen.de}

\author[1]{\fnm{Tim} \sur{Friede}}
\email{tim.friede@med.uni-goettingen.de}

\affil*[1]{
\orgdiv{Department of Medical Statistics},
\orgname{University Medical Center G\"ottingen},
\orgaddress{
\street{Humboldtallee~32},
\postcode{37073}~\city{G\"ottingen},
\country{Germany}
}
}

\affil[2]{
\orgdiv{Department of Cardiology and Pneumology},
\orgname{University Medical Center G\"ottingen},
\orgaddress{
\street{Robert-Koch-Stra{\ss}e~40},
\postcode{37075}~\city{G\"ottingen},
\country{Germany}
}
}

\affil[3]{
\orgname{Lung Clinic Immenhausen},
\orgaddress{
\street{Robert-Koch-Stra{\ss}e~3},
\postcode{34376}~\city{Immenhausen},
\country{Germany}
}
}

\affil[4]{
\orgdiv{Department of Pulmonary Medicine},
\orgname{University Medicine Essen --- Ruhrlandklinik,
University of Duisburg-Essen},
\orgaddress{
\street{T\"uschener Weg~40},
\postcode{45239}~\city{Essen},
\country{Germany}
}
}


\abstract{The abstract serves both as a general introduction to the topic and as a brief, non-technical summary of the main results and their implications. Authors are advised to check the author instructions for the journal they are submitting to for word limits and if structural elements like subheadings, citations, or equations are permitted.}

\abstract{
\textbf{Purpose:} Relative changes in event rates are a common primary endpoint in confirmatory asthma trials, and reliable assumptions about baseline event rates and extra-Poisson variability are essential for valid sample size planning. Using an example application in asthma, we aim to synthesize placebo-group exacerbation data, evaluate calendar-time trends, and assess implications for trial design.

\textbf{Methods:} 
We conducted a systematic review of randomized controlled trials (RCTs) reporting recurrent asthma exacerbations and extracted outcomes exclusively from their control groups, which were classified according to placebo type or asthma severity. Control-group outcomes were synthesized using hierarchical negative binomial models that accommodated heterogeneous reporting formats and jointly modeled event rates and overdispersion. To account for differences among control groups defined by placebo type or asthma severity, we considered two alternative approaches: separate models by subgroups or adjustment via covariates. Meta-regression was used to adjust for calendar-time trends. Trial design implications are illustrated by considering the example case of the subsequent NAVIGATOR trial.

\textbf{Results:} Sixty-two trials met the inclusion criteria. The overall analysis indicated an estimated relative change in annualized exacerbation rates of $-1.8\%$ (95\% CrI: $-7.5\%, 4.3\%$) per year, corresponding to an estimated $16.5\%$ decrease per decade (95\% CrI: $-54.2\%, 52.2\%$). Substantial between-study heterogeneity in both event rates and overdispersion was observed. (Within-trial) overdispersion~($\varphi$) was estimated as $1.53$ (95\% CrI: $0.60, 2.83$). In an illustrative phase~III design scenario, sample size requirements were sensitive to both the target subgroup and the uncertainty in the predicted event rate and overdispersion parameter.

\textbf{Conclusion:} 
Event rates in clinical trials exhibit considerable heterogeneity and potential temporal variation, and overdispersion represents a critical design parameter. Comprehensive modelling of recurrent-event endpoints allows for evidence-based and robust sample size calculations and trial design.
}

\keywords{Recurrent events; Negative binomial distribution; Overdispersion; Bayesian hierarchical model; Meta-regression; Sample size planning; Asthma trials}


\maketitle

\section{Introduction}\label{sec:intro}
Recurrent event endpoints are central to the evaluation of many chronic diseases. In conditions such as multiple sclerosis, chronic obstructive pulmonary disease (COPD), heart failure, and asthma, the frequencies of clinically relevant events---such as relapses, exacerbations or hospitalizations---serve as primary outcomes in confirmatory trials. Because such endpoints are typically analyzed using count-based regression models, reliable assumptions regarding baseline event rates and variability are essential for valid study design and interpretation. Asthma provides a particularly relevant example, as recurrent exacerbations constitute a defining feature of disease severity and long-term burden. Asthma continues to impose a substantial clinical and socio\-economic impact despite considerable advances in pharmacologic treatment. Severe exacerbations, defined by episodes of worsening respiratory symptoms requiring systemic corticosteroids, emergency department visits, or hospitalization, are associated with impaired lung function, reduced quality of life, increased health care utilization, and elevated mortality risk. Consequently, prevention of exacerbations represents a primary objective of long-term asthma management.

Reflecting their clinical relevance, exacerbation rates are frequently chosen as the primary endpoint in confirmatory phase~III trials in asthma. Because such endpoints involve recurrent events over a defined follow-up period, analyses are commonly performed using Poisson or negative binomial regression models \citep{Hilbe2011}. As emphasized by Keene et al.\ \cite{KeeneEtAl2007}, sample size requirements for recurrent-event trials depend not only on the assumed group-specific event rates but also on the extent of extra-Poisson variability (overdispersion). Failure to account for overdispersion may lead to underpowered studies or inflated type~I error rates when overdispersion is substantial.

Beyond statistical considerations at a single time point, accumulating evidence from other chronic diseases suggests that baseline event rates in RCTs may change over calendar time. Declining relapse rates have been documented in multiple sclerosis \cite{NicholasEtAl2011,SteinvorthEtAl2013,NicholasEtAl2019}, and similar temporal patterns have been reported in chronic obstructive pulmonary disease and heart failure \cite{AndreasEtAl2019,AndreasEtAl2022,KimmounEtAl2021}. Such trends may arise from improved background therapy, broader implementation of preventive measures, earlier diagnosis, changes in eligibility criteria, or evolving patient selection processes. If baseline exacerbation rates evolve over time but trial designs continue to rely on historical assumptions, sample size calculations may no longer be adequate. In asthma, however, systematic evidence regarding potential temporal trends in exacerbation rates remains limited. In particular, it remains unclear whether exacerbation rates have changed over the past decades across placebo-type or asthma-severity groups in RCTs, and how such differences might affect planning assumptions for future studies. Moreover, overdispersion in event counts has not been comprehensively synthesized across trials.

Temporal changes may hence affect both the comparability across trials, as well as the assumptions used in sample size calculation and power determination~\cite{FriedePohlmannSchmidli2019}. In the present work, we focus on asthma as a clinical application and conduct a systematic review of RCTs with recurrent exacerbation endpoints to investigate potential calendar-time trends in exacerbation rates across placebo-type and asthma-severity groups. We analyse the aggregated recurrent-event data using hierarchical models, considering both separate analyses by subgroups as well as covariate-based models that incorporate placebo type or asthma severity as study-level covariates in the event-rate model. We quantify between-study heterogeneity in event rates and overdispersion parameters and evaluate the implications of these findings for sample size determination in future confirmatory trials.

The remainder of this article is structured as follows.
Section~\ref{sec:MAandSR} introduces a motivating examples, hightlighting the challenges faced at the trial design stage, and describes the literature review conducted to form an empirical data base for informed design choices.
In Section~\ref{sec:methods}, we outline the background and methodological motivation for modeling recurrent event endpoints and addressing overdispersion in confirmatory trials. This section first introduces the proposed statistical framework, including the negative binomial model, likelihood construction accommodating heterogeneous reporting formats, modeling of calendar-time trends, and sample size planning based on event rates and dispersion parameters.
Section~\ref{sec:results} presents the application of the modeling framework to the collected dataset, including estimation of temporal trends and between-study heterogeneity, as well as implications for trial design. Finally, the paper concludes in Section~\ref{sec:discussion} with a discussion of methodological considerations, limitations, and directions for future methodological research.

\section{Motivating example and systematic review}\label{sec:MAandSR}
\subsection{The motivating example: design of an asthma trial}\label{sec:AsthmaExample}
To illustrate the implications of baseline event-rate and overdispersion
assumptions for trial design, we consider the phase~III \textsc{navigator}
trial, a randomized, double-blind, placebo-controlled, two-arm study evaluating
\emph{tezepelumab} in patients with severe, uncontrolled asthma
\cite{menzies2021tezepelumab,Navigator}.
The trial enrolled adolescents and adults aged 12--80 years. Eligible patients
were required to have uncontrolled asthma, reduced baseline lung function, and
at least two asthma exacerbations during the preceding year despite treatment
with medium- or high-dose inhaled corticosteroids (ICS) plus at least one
additional controller medication, with or without oral corticosteroids.
Patients were randomized in a 1:1 ratio to receive \emph{tezepelumab} or placebo
every four weeks over a 52-week treatment period, while continuing their
prescribed background therapy. The primary endpoint was the annualized asthma
exacerbation rate. The total sample size was approximately 1060~patients
(530~patients per group), within a confirmatory phase~III framework
incorporating hierarchical testing, multiplicity control, and key secondary
endpoints.

The \textsc{navigator} design was based on an assumed annualized exacerbation
rate of $0.9$~events per year in the placebo group, a negative binomial shape
parameter of~2.4 (corresponding to $\varphi=1/2.4=0.417$ in the
parameterization introduced in Section~\ref{sec:NB} below), and a 10\% dropout
rate. Under these assumptions, the protocol reported at least 99\% power to
detect a 50\% rate reduction at a two-sided 1\% significance level. The
relatively stringent significance and power levels were intended to yield
\emph{statistically persuasive} evidence. In contrast, our calculations are not
meant to reproduce the original \textsc{navigator} setup; instead, we aim for a
conventional target power of~80\% and a two-sided significance level of~5\% to
illustrate sample size requirements under common planning assumptions. For the
treatment effect, we again assume a 50\% reduction in the event rate as
clinically relevant.

The \textsc{navigator} population is particular in that it included both adults and adolecents in the study population, and that it focused on cases of \emph{severe} asthma. To provide an empirical basis for alternative planning assumptions, we
therefore consider control-group data from a systematic review of historical asthma trials. The synthesis of control-group event rates and overdispersion, while accounting for placebo type, asthma severity, and calendar time, provides
a range of historically informed assumptions for illustrating their implications for sample size planning.

\subsection{The literature review}\label{sec:LitReview}
To provide an empirical basis for the study design setup,
we performed a systematic literature search to identify placebo-controlled asthma trials reporting exacerbation outcomes. We queried the \emph{PubMed} and \emph{Cochrane CENTRAL} data bases using the search terms \textquotedblleft{asthma* AND (double-blind OR triple-blind) AND (exacerba* OR hospitali*)}\textquotedblright, using the limits \textquotedblleft{Clinical Trial}\textquotedblright, \textquotedblleft{Randomized Controlled Trial}\textquotedblright and \textquotedblleft{adults}\textquotedblright. In order to also retrieve studies that have only recently been added to the database and may not be completely indexed, we repeated the search using only the limit \textquotedblleft{up to present}\textquotedblright.  The date of last search was January~2nd, 2019. The list of abstracts was independently reviewed by three authors (CR, SS, JH) using the following inclusion criteria: (i)~the study needs to deal with adult Asthma patients; (ii)~the study has a placebo control group; (iii)~the study design uses parallel group or a cross-over design; (iv)~the trial is double-blind and randomized; (v)~exacerbation rates are quoted or can be calculated from the data presented; (vi)~at least 100 patients (intention-to-treat population) are included into the study; (vii)~the study has a treatment duration of at least 12~weeks.

\subsection{Data extraction}\label{sec:DataExtraction}
We recorded information on patient demographics, trial eligibility criteria, exacerbation rates, and the corresponding uncertainties as well as overdispersion. Information on studies was assembled in a table and checked independently by at least two authors (CR; SS; SA; HW; JH). The reporting of rates was rather heterogeneous, partly because not all studies considered here treated exacerbation rates as a primary endpoint \cite{AaronEtAl2008}. The preferred source of evidence on rates are explicitly quoted rate estimates, and corresponding standard errors or confidence intervals. If a rate is given along with the number of exacerbation-free patients, this pair of figures provides some evidence on the overdispersion. If only one of the numbers is given, these may still be used to fit the joint model \cite{RoeverAndreasFriede2016}. Quoted rates or total exacerbation counts may be converted to one another, as can numbers and fractions of (non-)exacerbating patients. When exacerbation counts of differing severities are given, the numbers pertaining to ``moderate to severe'' exacerbation events were used. Our review focuses on data from the studies’ placebo groups. Only the concurrent use of inhaled corticosteroids (ICS),  long-acting beta agonist (LABA) or rescue medications was acceptable in order to be considered as placebo groups for purposes of our analysis. We distinguished between studies involving ``true placebo'', ``ICS-placebo'' and ``ICS-LABA-placebo'' groups where ``true'' placebo indicates that ICS and LABA were not allowed while in the ``ICS'' and ``ICS-LABA'' groups, the corresponding co-medications were allowed. In addition, we considered asthma severity levels. Classification was based on the mean baseline forced expiratory volume in 1~second (\fevone, in litres). {\fevonepp} is defined as the \fevone expressed as a percentage of its predicted value. As per the 2020 NAEPP guidelines \cite{cloutier20202020}, asthma severity can be categorized using the following criteria: Mild Asthma: $\mbox{\fevonepp} \geq 80 \%$; Moderate Asthma: $60\% \leq \mbox{\fevonepp} < 80\%$; Severe Asthma: $\mbox{\fevonepp} < 60\%$. Other extracted study characteristics were: number of participants, study duration, number of patient-years, fraction of smokers, mean pack-years,  mean Asthma Quality-of-Life Questionnaire score (AQLQ), mean Asthma Control Test score (ACT) and mean eosinophil count per microlitre. Study quality was assessed by two reviewers (SS; SA) using the Oxford Quality Scale score \cite{JadadEtAl1996}; discrepancies in this regard were resolved by a third reviewer (TF).

\section{Methods}\label{sec:methods}
We begin by formally introducing the negative binomial  model for recurrent event data (Section~\ref{sec:NB}), followed by the trial design considerations (the sample size formula and its dependence on event rates and overdispersion; Section~\ref{trialdesign}). Subsequently, we present the hierarchical synthesis framework for aggregated recurrent event outcomes, including the construction of a joint likelihood for heterogeneous reporting formats. Finally, we extend this framework to meta-regression models that incorporate calendar time and subgroup characteristics as study-level covariates (Section~\ref{aggregated}).

\subsection{The negative binomial model}\label{sec:NB}
Modeling of event count data here is based on the negative binomial distribution, which generalizes the Poisson model. The Poisson distribution is appropriate for modeling event counts that occur at a constant rate over time \cite{SpiegelhalterEtAl}. If a patient experiences events at an annualized rate~$\lambda$ and is observed for a duration of $d$~years, the total number of observed events follows a Poisson distribution with parameter $d\lambda$. Under this model,
\begin{equation}
  \expect[X] = d\lambda 
  \quad \text{and} \quad 
  \var(X) = d\lambda,
\end{equation}
so that the mean and variance coincide. This equidispersion property is a defining characteristic of the Poisson model. In many clinical applications, however, observed event counts exhibit greater variability than permitted by the Poisson assumption \cite{AaronEtAl2008,EMA2012,EMA2015b,AndreasEtAl2019}.
Such extra-Poisson variability may arise from heterogeneity between patients, for example due to differing baseline risks or disease activity levels. The negative binomial distribution extends the Poisson model by introducing an overdispersion parameter~$\varphi > 0$, allowing the variance to exceed the mean. A common motivation for the negative binomial model is as a Poisson mixture in which the individual rate parameter itself is varying. In this formulation, patients are assumed to have similar but not identical underlying event rates, and the resulting marginal distribution of event counts follows a negative binomial distribution.
Under the negative binomial model, the expected number of events 
remains unchanged, i.e.
\begin{equation}
  \expect[X] = d\lambda,
\end{equation}
while the variance becomes
\begin{equation}
  \var(X) = d\lambda \bigl(1 + \varphi d\lambda \bigr),
\end{equation}
which exceeds the mean whenever~$\varphi > 0$. Thus, the parameter $\varphi$ directly quantifies the degree of extra-Poisson variability. 
In the limiting case $\varphi \to 0$, the  negative binomial model again reduces to the Poisson model.

\subsection{Sample size calculation}\label{trialdesign}
Sample size determination for clinical trials involving overdispersed count outcomes requires specification of two key nuisance parameters: the control group event rate and the overdispersion parameter. Misspecification of these parameters may result in studies that are either under- or overpowered, both of which are undesirable for ethical and economical reasons \cite{kim2021handbook}.

Consider a two-arm trial with arms indexed by $i\in\{1,2\}$ and $n_i$ patients in arm~$i$. Let $\lambda_1$ denote the baseline (control-group) event rate, and let $\lambda_2$ denote the event rate under the experimental treatment. The treatment effect is expressed through the logarithmic rate ratio
\begin{equation}
  \delta_1 = \log(\lambda_1) - \log(\lambda_2) = \log(\lambda_1 / \lambda_2);
\end{equation}
at the trial design stage, a \emph{minimal clinically important difference (MCID)} is commonly considered here.

To derive the arm-level variance required for the subsequent sample size calculation, we adopt an individual-level negative binomial working model. Let $X_{ij}$ denote the number of recurrent events observed for patient $j=1,\dots,n_i$ in trial arm~$i$ over follow-up time $d_{ij}$. Conditional on $d_{ij}$, the event counts are assumed to follow a
negative binomial distribution with mean $\mu_{ij}=\lambda_i d_{ij}$ and overdispersion parameter~$\varphi$, implying
\begin{equation}
  \var(X_{ij})
  =
  \mu_{ij}\bigl(1+\varphi\mu_{ij}\bigr).
\end{equation}

We assume a treatment-to-control allocation ratio of $1:r$, with $n$ patients in the treatment group and $r n$~patients in the control group.
Assuming equal follow-up $d_{ij}=d$ for all patients, we test
\begin{equation}
  H_0:\;\delta_1=\log(\lambda_1)-\log(\lambda_2)=0,
\end{equation}
or equivalently,
\begin{equation}
  H_0:\;\lambda_1/\lambda_2=1.
\end{equation}
To achieve a two-sided type~I error rate~$\alpha$ and power~$1-\beta$
under the alternative $H_1:\;\delta_1\neq0$, the required sample size in the
treatment group is
\begin{equation}\label{eqn:NBSampleSize}
n \;=\;
\frac{(z_{1-\alpha/2}+z_{1-\beta})^2}{\delta_1^2}
\left\{
\frac{1}{r}
\left(
\frac{1}{d\lambda_1}+\varphi
\right)
+
\left(
\frac{1}{d\lambda_2}+\varphi
\right)
\right\},
\end{equation}
and the corresponding control-group sample size is $rn$ \cite{chow2017sample}.

\subsection{Synthesis of aggregated recurrent event outcomes}\label{aggregated}
\subsubsection{The hierarchical meta-regression model}
Count data commonly arise as endpoints in clinical trials, however, recurrent event outcomes are often reported using heterogeneous summary measures across studies, such as (i)~rates with standard errors, (ii)~total event counts, (iii)~proportions of event-free patients, or (iv)~combinations thereof. These reporting formats can be viewed as alternative summaries relating to a common underlying event-count process.
To accommodate such heterogeneous data, we employ a Bayesian hierarchical modeling framework based on a negative binomial (NB) data-generating mechanism.
A joint likelihood is constructed that links the different reporting formats to shared latent study-specific parameters, namely, the event rate~($\lambda$) and the overdispersion parameter~($\varphi$). This formulation allows all available information to contribute coherently to the estimation of rate and dispersion parameters.
Technical details of the likelihood construction for the different
reporting formats are provided in Appendix~\ref{AppendixA} and in
\cite{RoeverAndreasFriede2016,AndreasEtAl2019}.

For study~$i=1,\dots,I$ with follow-up duration~$d_i$, the placebo-group event counts are assumed to follow a negative binomial distribution with mean~$d_i\lambda_i$ and overdispersion parameter~$\varphi_i$.
Between-study variability in both the baseline event rates and the overdispersion parameters is modeled hierarchically within a Bayesian framework.
The study-specific rate parameters may depend on study-level covariates through a meta-regression structure while retaining residual heterogeneity across studies. Specifically, the general
event-rate model is written as
\begin{equation} \label{eqn:rateModel}
\log(\lambda_i) \sim
\normaldistn(\mathbf{x}_i^\top\boldsymbol{\beta}, \sigma_\lambda^2),
\end{equation}
where~$\mathbf{x}_i$ denotes a vector of study-level covariates and
$\boldsymbol{\beta}$ denotes the corresponding regression coefficients. For
example, $\mathbf{x}_i$ may include calendar time, baseline disease severity,
placebo type, treatment background, or other study-level design characteristics.
The study-specific overdispersion parameters were modeled using a random effects model as
\begin{equation} \label{eqn:OverdispModel}
\log(\varphi_i) \sim
\normaldistn\bigl(\mu_\varphi,\,\sigma_\varphi^2\bigr),
\end{equation}
and with that allowing for between-study heterogeneity in the extra-Poisson variability.

\subsubsection{Prior specification}
For the analysis of the assembled asthma RCT data, placebo type and asthma severity were considered as study-level grouping variables, while calendar time was included as a continuous covariate. Priors for intercept and slope were specified as
\begin{equation}
  \beta_0 \sim \unifdistn\bigl(\log(10^{-3}), \, \log(10^3)\bigr),
  \qquad
  \beta_1 \sim \normaldistn\bigl(0, \, 10^2\bigr),
\end{equation}
essentially mimicking an uninformative (uniform) prior.
Priors for the random effects' parameters were set to 
\begin{equation}
  \sigma_\lambda \sim \halfnormaldistn(1.0),
  \quad
  \mu_\varphi \sim \unifdistn\bigl(\log(10^{-4}), \log(10^4)\bigr),
  \quad
  \sigma_\varphi \sim \halfnormaldistn(1.0).
\end{equation}
For covariate-based models with group-specific intercepts and covariate effects,
the same priors were assigned independently to each~$\beta_{0,g}$ and
$\beta_{1,g}$.
The choice of prior distributions was motivated by previous work in
hierarchical modeling of recurrent event data~\cite{RoeverAndreasFriede2016}.
The models were fitted using Markov chain Monte Carlo (MCMC) methods implemented
in JAGS via the~\texttt{rjags} package in R~\cite{R:rjags}.
Posterior inference is summarized using posterior medians and central 95\% credible intervals (CrIs).

\subsubsection{Additional analyses}
As an additional exploratory analysis, we examined associations among
study-level characteristics, including disease severity and other clinical or
design covariates, using mixed-effects meta-regression models. This analysis was
conducted to assess potential correlations between study characteristics and to
identify possible confounding structures that could inform model specification.
Stratified meta-regression analyses were also performed to explore
subgroup-specific patterns across different study-level classifications. These
auxiliary analyses were conducted using the~\texttt{metafor} R
package~\cite{Viechtbauer2010}.

\subsubsection{Posterior predictive assurance}
The sample size formula~(\ref{eqn:NBSampleSize}) is based on \emph{fixed}
values of the event rates and overdispersion parameter. To account for
uncertainty in these planning parameters, we consider the Bayesian
\emph{assurance} framework~
\cite{stallard1998sample,mutze2018sample,grieve2022hybrid}.
Let $\pi(n;\delta_1,\boldsymbol{\theta},\mathbf{x})$ denote the frequentist
power of a planned trial with $n$ patients in the treatment group and $rn$
patients in the control group, corresponding to a treatment-to-control
allocation ratio of $1:r$. The power is conditional on the prespecified target
log rate ratio~$\delta_1$, the uncertain planning
parameters~$\boldsymbol{\theta}$, and the prespecified trial
characteristics~$\mathbf{x}$. Given a planning distribution
$p(\boldsymbol{\theta}\mid\mathbf{x})$, assurance is defined as

\begin{equation}
\assurance(n,\delta_1,\mathbf{x})
=
\expect_{\boldsymbol{\theta}\mid\mathbf{x}}
\bigl[
\pi(n;\delta_1,\boldsymbol{\theta},\mathbf{x})
\bigr]
=
\int
\pi(n;\delta_1,\boldsymbol{\theta},\mathbf{x})
\, p(\boldsymbol{\theta}\mid\mathbf{x})
\, \differential\boldsymbol{\theta}\mbox{,}
\end{equation}
that is, the expected power, marginalized over uncertainty in the planning
parameters.

In our application, the uncertain planning parameters are
$\boldsymbol{\theta}=(\lambda_1,\varphi)$, comprising the control-group event
rate~$\lambda_1$ and overdispersion parameter~$\varphi$ in a future trial,
while~$\mathbf{x}$ contains the relevant study-level characteristics. Their
posterior predictive distributions are induced by the hierarchical models in
Equations~(\ref{eqn:rateModel}) and~(\ref{eqn:OverdispModel}). Prediction of
the event rate accounts for the regression parameters and residual
between-study heterogeneity~$\sigma_\lambda$, whereas prediction of the
overdispersion parameter accounts for both the central log-overdispersion
level~$\mu_\varphi$ and between-study heterogeneity~$\sigma_\varphi$.
Consequently, the planning distribution incorporates both posterior
uncertainty in the model parameters and predictive study-level variation in
the event rate and overdispersion parameter.

For a prespecified target log rate ratio
$\delta_1=\log(\lambda_1/\lambda_2)$, the treatment-group event rate is
\[
\lambda_2=\lambda_1\exp(-\delta_1).
\]
Using the notation introduced in Section~\ref{trialdesign}, the large-sample
variance of the estimated log rate ratio~$\hat{\delta}_1$ can be written as
\begin{equation}
\textstyle
\var\bigl(
\hat{\delta}_1
\mid\delta_1,\boldsymbol{\theta},\mathbf{x}
\bigr)
=
\frac{V(\delta_1,\boldsymbol{\theta},\mathbf{x})}{n},
\quad
\mbox{where}
\quad
V(\delta_1,\boldsymbol{\theta},\mathbf{x})
=
\frac{1}{r}
\left(
\frac{1}{d\lambda_1}+\varphi
\right)
+
\left(
\frac{1}{d\lambda_2}+\varphi
\right)
\end{equation}
\cite{friede2010ablinded,friede2010bblinded}.
Defining
\begin{equation}
\textstyle
\eta(n;\delta_1,\boldsymbol{\theta},\mathbf{x})
=
\sqrt{
\frac{n\delta_1^2}
     {V(\delta_1,\boldsymbol{\theta},\mathbf{x})}
},
\end{equation}
the conditional power of the two-sided Wald test is
\begin{equation}
\pi(n;\delta_1,\boldsymbol{\theta},\mathbf{x})
=
\Phi\left(
\eta(n;\delta_1,\boldsymbol{\theta},\mathbf{x})
-z_{1-\alpha/2}
\right)
+
\Phi\left(
-\eta(n;\delta_1,\boldsymbol{\theta},\mathbf{x})
-z_{1-\alpha/2}
\right),
\end{equation}
where $\Phi(\cdot)$ denotes the standard normal cumulative distribution
function. For given historical data~$\histdata$, the posterior predictive
assurance is therefore
\begin{equation}
\assurance(n,\delta_1,\mathbf{x};\histdata)
=
\int
\pi(n;\delta_1,\boldsymbol{\theta},\mathbf{x})
\,p(\boldsymbol{\theta}\mid\mathbf{x},\histdata)
\,\differential\boldsymbol{\theta}.
\end{equation}
For a target assurance level~$\gamma$, the required treatment-group sample
size then is
\begin{equation}
n_\gamma(\delta_1,\mathbf{x};\histdata)
=
\min\left\{
n\in\mathbb{N}:\,
\assurance(n,\delta_1,\mathbf{x};\histdata)\geq\gamma
\right\}\mbox{,}
\end{equation}
and the corresponding total sample size is
\[
N_\gamma=(1+r)n_\gamma\mbox{.}
\]
The posterior predictive expectations were approximated using draws generated
from the MCMC output of the hierarchical model.
\section{Results}\label{sec:results}
\subsection{Studies identified}\label{sec:Studies identified}
According to the predefined eligibility criteria, a total of 2,725~records were identified, including 514 from PubMed, 489 from Embase, and 1,722 from Cochrane CENTRAL. After removal of 1,020 duplicate records, 1,705~records remained for title and abstract screening. Eventually, 538~reports underwent full-text assessment. Reports were excluded at the full-text stage according to the inclusion criteria, including insufficient sample size or study duration, ineligible populations or study designs, absence of relevant exacerbation outcomes or placebo groups, and inappropriate background therapy
(see Sections~\ref{sec:LitReview} and~\ref{sec:DataExtraction}).

Ultimately, 62 studies were included in the meta-analysis; some publications contributed more than one eligible study. The complete search and screening process, including detailed reasons for exclusion, is summarized in the PRISMA flow diagram in Appendix~\ref{sec:AppendixB} (Figure~\ref{fig:PrismaFlowChart}).

\begin{table}[!htbp]
\centering
\small
\caption{Characteristics of the included studies. Values are presented as
medians and ranges across studies; $N$ denotes the number of studies with
available information.}
\label{table:baseline}

\begin{tabular}{@{}lccc@{}}
\toprule
Characteristic & $N$ & Median & Range \\
\midrule
No. of patients
& 62 & 176 & 27--3568 \\

Study duration (yr)
& 62 & 0.537 & 0.067--3 \\

Males (\%)
& 59 & 38.2 & 25.9--61.65 \\

Mean age (yr)
& 55 & 45.7 & 24.0--55.0 \\

Mean \fevone{} (L)
& 40 & 2.125 & 1.558--3.389 \\

Mean \fevonepp{} (\%)
& 57 & 68.0 & 52.69--96.4 \\

Smokers (\%)
& 28 & 3.95 & 0--38 \\

Mean AQLQ
& 27 & 4.22 & 3.60--5.50 \\

Mean ec ($\mu$L$^{-1}$)
& 20 & 337.5 & 0.25--688 \\
\bottomrule
\end{tabular}
\end{table}

A total of 62~studies were included in the analysis; their study characteristics are summarized in Table~\ref{table:baseline}. The median sample size was 176~participants (range 27--3568), and the median study duration was 0.54~years (range 0.07--3~years). Across all studies, the median proportion of male participants was 38.2\% (range 25.9\%--61.7\%), and the median mean age was 45.7~years (range 24--55~years).
The median baseline forced expiratory volume in 1 second (\fevone), expressed as percentage of the predicted value (\fevonepp), was 68.0\% (range 52.7\%--96.4\%). The median {\fevone} was 2.13~L (range 1.56--3.39~L). The median proportion of smokers was 4.0\% (range 0--38\%). The median Asthma Quality of Life Questionnaire (AQLQ) score was 4.22 (range 3.6--5.5). The median eosinophil count was 337.5~$\mu$L$^{-1}$ (range 0.25--688).
Among the included studies, 27 used true placebos, 16 used ICS-placebo, and 19 used ICS-LABA-placebo. Sample size and study duration varied across subgroups, with ICS-LABA-placebo studies generally having larger sample sizes (median 234~participants) and longer duration (median 0.92~years). Baseline demographic and lung function characteristics were broadly comparable across subgroups. Detailed study characteristics according to placebo type and
asthma severity are provided in Table~\ref{table:baseline_subgroups} in Appendix~\ref{AppendixE}. 

\subsection{Trends in exacerbation rates}\label{sec:exacerbation-trends}
To assess calendar-time trends, publication year (centered at the year~2000) was included as a study-level covariate in the event-rate model. The overall model included all 62~studies and estimated a common intercept and time trend. Potential differences according to placebo type and asthma severity were examined using two complementary strategies. In the subgroup-specific analyses, the hierarchical model was fitted independently to each data subset, whereas the covariate-based analyses included all studies jointly using group indicators and their interactions with publication year. The latter approach allowed group-specific intercepts and time trends while sharing the heterogeneity and overdispersion parameters across groups. 

Data on exacerbations were reported in heterogeneous formats.
Nine studies provided rate estimates with corresponding standard errors, 17 studies reported total exacerbation counts together with the number of exacerbation-free patients, 13 studies reported only total counts, and 23 studies provided only the proportion of non-exacerbating patients.

\begin{figure}
    \centering
    \includegraphics[width=0.95\linewidth]{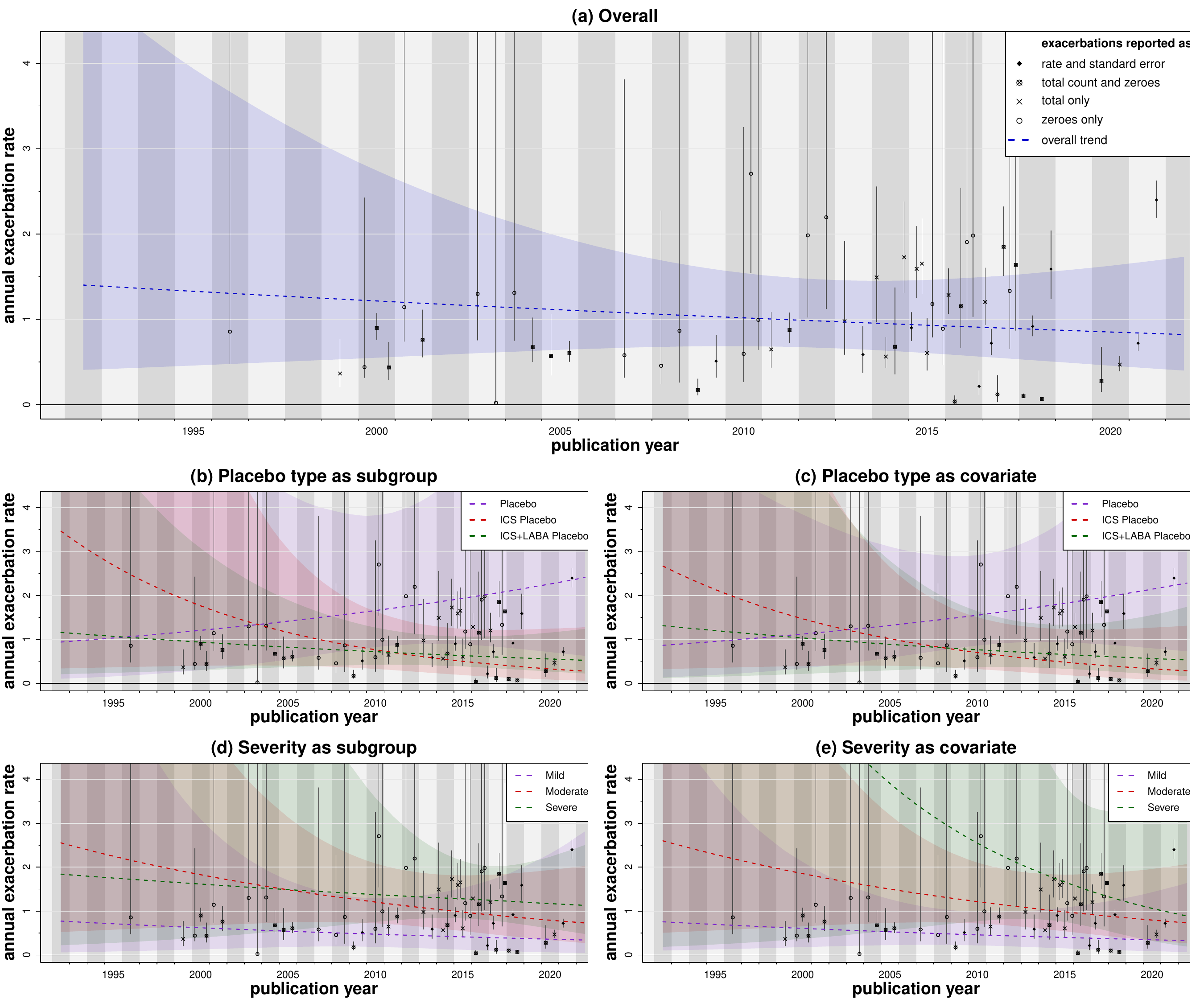}
    \caption{Time trends in annual exacerbation rates across the overall evidence base and according to placebo type or asthma severity. Panel~(a) shows the overall trend across all studies; panels~(b) and~(c) show analyses according to placebo type using independent subgroup-specific and joint covariate-based models, respectively. Panels~(d) and~(e) show the corresponding analyses according to asthma severity. Dashed lines show the estimated annual exacerbation rates over publication year, with shaded areas representing 95\% credible regions. Points and vertical whiskers represent study-specific model-based estimates and their 95\% credible intervals. The shown (``shrinkage'') estimates are based not only on the corresponding study’s provided data (which in some cases may not be sufficient to derive a rate estimate), but these are supported by the complete data set. Different types of data
    sources are indicated by different symbols (some references directly provide rate estimates along with standard errors, some report the total number of exacerbations and the number of exacerbation-free patients, and some only one of the two).}
    \label{fig:AsthmaTrendPlot}
\end{figure}

In the overall analysis of all 62 studies, the estimated association between publication year and exacerbation rate corresponded to a relative change of $-1.8$\% per year (95\% CrI: $-7.5$\%, 4.3\%), equivalent to an estimated 16.5\% decrease per decade (95\% CrI: $-54.2$\%, 52.2\%). The overall trend is shown in Figure~\ref{fig:AsthmaTrendPlot}(a). The placebo-type analyses are shown in Figure~\ref{fig:AsthmaTrendPlot}(b) and~\ref{fig:AsthmaTrendPlot}(c) for the subgroup-specific and covariate-based models, respectively, and the corresponding analyses by baseline severity are shown in Figure~\ref{fig:AsthmaTrendPlot}(d) and~\ref{fig:AsthmaTrendPlot}(e).
Detailed parameter estimates are provided in Table~\ref{tab:regression-ana-all}.

When placebo type was considered, temporal trends varied in direction but remained imprecisely estimated. In the subgroup-specific analyses, studies using true placebos showed an estimated annual relative change of 3.2\% (95\% CrI: $-7.9$\%, 15.3\%), whereas ICS-placebo studies showed an annual relative change of $-8.1$\% (95\% CrI: $-17.5$\%, 3.3\%) and ICS--LABA-placebo studies an annual relative change of $-2.6$\% (95\% CrI: $-10.2$\%, 5.4\%). The corresponding covariate-based analyses gave broadly similar conclusions: the annual relative changes were estimated as 
3.3\% (95\% CrI: $-5.9$\%, 13.3\%) for true-placebo studies, $-7.2$\% (95\% CrI: $-16.8$\%, 3.6\%) for ICS-placebo studies, and $-3.0$\% (95\% CrI: $-13.2$\%, 8.2\%) for ICS--LABA-placebo studies.

Analyses according to asthma severity also showed imprecise and heterogeneous temporal patterns. In the subgroup-specific analyses, the annual relative changes were estimated as $-2.7$\% (95\% CrI: $-15.3$\%, 12.1\%) for mild asthma, $-4.1$\% (95\% CrI: $-11.6$\%, 3.8\%) for moderate asthma, and $-1.6$\% (95\% CrI: $-13.3$\%, 7.8\%) for severe asthma. In the covariate-based analyses, the corresponding estimates were $-2.7$\% (95\% CrI: $-12.8$\%, 8.4\%), $-4.2$\% (95\% CrI: $-11.7$\%, 4.0\%), and $-8.3$\% (95\% CrI: $-25.9$\%, 9.5\%), respectively.

Neither the overall analysis nor the subgroup analyses provided convincing evidence of a calendar-time trend in exacerbation rates, as the credible intervals included a zero slope. The subgroup estimates were generally more uncertain because they were based on fewer studies or involved additional parameters. Apparent differences across placebo-type or severity groups and between modelling strategies should therefore not be over-interpreted.

Although neither the overall nor the subgroup analyses provided clear evidence
of a calendar-time trend, the estimated overdispersion parameters suggested
extra-Poisson variability in patient-level event counts within studies. In the
overall analysis, the estimated central log-overdispersion parameter was
$\mu_\varphi=0.422$ (95\% CrI: $-0.503$, 1.041). In the
subgroup-specific analyses, the corresponding estimates ranged from 0.232 to
0.854 across placebo-type subgroups and from $-3.650$ to 0.507 across
asthma-severity subgroups. The particularly low estimate for the severe-asthma
subgroup, $\mu_\varphi=-3.650$ (95\% CrI: $-8.933$, 0.999), was accompanied
by substantial uncertainty and should therefore be interpreted cautiously. In
the covariate-based analyses, the estimated common central
log-overdispersion parameters were $\mu_\varphi=0.498$ for the placebo-type
model and $\mu_\varphi=0.465$ for the asthma-severity model.

Beyond the within-study extra-Poisson variability captured by
$\varphi_i$, substantial between-study heterogeneity was observed in both the
underlying negative binomial event rates and the extent of overdispersion. In
the overall analysis, the estimated between-study standard deviations on the
log scale were $\sigma_{\lambda}=1.419$ (95\% CrI: 1.136, 1.784) for the
event rates and $\sigma_{\varphi}=1.226$ (95\% CrI: 0.722, 2.058) for the
overdispersion parameters. In the subgroup-specific analyses,
$\sigma_{\lambda}$ ranged from 0.979 to 1.628 across placebo-type subgroups
and from 0.726 to 1.752 across asthma-severity subgroups, while
$\sigma_{\varphi}$ ranged from 0.804 to 1.281 and from 0.686 to 1.289,
respectively. In the covariate-based analyses, the corresponding common
heterogeneity estimates were $\sigma_{\lambda}=1.391$ (95\% CrI: 1.089,
1.777) and $\sigma_{\varphi}=1.214$ (95\% CrI: 0.700, 2.048) for the
placebo-type model, and $\sigma_{\lambda}=1.385$ (95\% CrI: 1.112, 1.749)
and $\sigma_{\varphi}=1.197$ (95\% CrI: 0.727, 1.969) for the
asthma-severity model.

These parameters describe conceptually distinct sources of variability. The
negative binomial overdispersion parameter~$\varphi_i$ captures extra-Poisson
variability in event counts within study~$i$, whereas
$\sigma_{\lambda}$ captures between-study heterogeneity in exacerbation rates
and $\sigma_{\varphi}$ captures between-study heterogeneity in the extent of
overdispersion. The hierarchical negative binomial framework accommodates
these sources of variability simultaneously when synthesizing exacerbation
outcomes reported in heterogeneous formats.

\begin{table}[htbp]
\footnotesize
\centering
\setlength{\tabcolsep}{2pt}
\renewcommand{\arraystretch}{0.82}
\caption{Parameter estimates and annual percentage changes from subgroup-specific and covariate-based regression analyses in the overall dataset, in the subgroups of true-placebo, ICS-placebo and ICS--LABA-placebo studies, and in studies of mild, moderate and severe asthma.}
\label{tab:regression-ana-all}

\begin{tabular}{@{}lcc@{\hspace{3pt}}cc@{}}
\toprule
parameter 
& \multicolumn{2}{c}{subgroup model} 
& \multicolumn{2}{c}{covariate model} \\
\cmidrule(lr){2-3} \cmidrule(lr){4-5}
& estimate (95\% CrI) 
& annual \% change 
& estimate (95\% CrI) 
& annual \% change \\
\midrule

\textbf{All studies (62 studies)} & & & & \\
rate intercept $\beta_0$ 
& 0.195 (-0.596, 1.009) 
&  
& 0.195 (-0.596, 1.009) 
&  \\
rate slope $\beta_1$ (publication year) 
& -0.018 (-0.078, 0.042) 
& -1.767 (-7.480, 4.295) 
& -0.018 (-0.078, 0.042) 
& -1.767 (-7.480, 4.295) \\
rate heterogeneity $\sigma_{\lambda}$ 
& 1.419 (1.136, 1.784) 
&  
& 1.419 (1.136, 1.784) 
&  \\
overdispersion $\mu_{\varphi}$ 
& 0.422 (-0.503, 1.041) 
&  
& 0.422 (-0.503, 1.041) 
&  \\
overdispersion heterogeneity $\sigma_{\varphi}$ 
& 1.226 (0.722, 2.058) 
&  
& 1.226 (0.722, 2.058) 
&  \\
\midrule

\textbf{True placebo (27 studies)} & & & & \\
rate intercept $\beta_0$ 
& 0.189 (-1.191, 1.765) 
&  
& 0.116 (-1.069, 1.346) 
&  \\
rate slope $\beta_1$ (publication year) 
& 0.031 (-0.082, 0.142) 
& 3.194 (-7.882, 15.261) 
& 0.032 (-0.061, 0.125) 
& 3.297 (-5.887, 13.338) \\
rate heterogeneity $\sigma_{\lambda}$ 
& 1.628 (1.176, 2.341) 
&  
& 1.391 (1.089, 1.777) 
&  \\
overdispersion $\mu_{\varphi}$ 
& 0.625 (-0.734, 1.499) 
&  
& 0.498 (-0.398, 1.102) 
&  \\
overdispersion heterogeneity $\sigma_{\varphi}$ 
& 1.275 (0.719, 2.230) 
&  
& 1.214 (0.700, 2.048) 
&  \\
\midrule

\textbf{ICS-placebo (16 studies)} & & & & \\
rate intercept $\beta_0$ 
& 0.568 (-0.913, 1.942) 
&  
& 0.385 (-0.933, 1.747) 
&  \\
rate slope $\beta_1$ (publication year) 
& -0.084 (-0.192, 0.032) 
& -8.078 (-17.505, 3.292) 
& -0.075 (-0.184, 0.036) 
& -7.211 (-16.833, 3.625) \\
rate heterogeneity $\sigma_{\lambda}$ 
& 1.328 (0.827, 2.054) 
&  
& 1.391 (1.089, 1.777) 
&  \\
overdispersion $\mu_{\varphi}$ 
& 0.854 (-0.799, 2.321) 
&  
& 0.498 (-0.398, 1.102) 
&  \\
overdispersion heterogeneity $\sigma_{\varphi}$ 
& 1.281 (0.167, 2.496) 
&  
& 1.214 (0.700, 2.048) 
&  \\
\midrule

\textbf{ICS--LABA-placebo (19 studies)} & & & & \\
rate intercept $\beta_0$ 
& -0.063 (-1.196, 1.123) 
&  
& 0.033 (-1.531, 1.662) 
&  \\
rate slope $\beta_1$ (publication year) 
& -0.026 (-0.107, 0.052) 
& -2.606 (-10.153, 5.377) 
& -0.030 (-0.141, 0.079) 
& -2.956 (-13.157, 8.227) \\
rate heterogeneity $\sigma_{\lambda}$ 
& 0.979 (0.703, 1.447) 
&  
& 1.391 (1.089, 1.777) 
&  \\
overdispersion $\mu_{\varphi}$ 
& 0.232 (-1.575, 1.014) 
&  
& 0.498 (-0.398, 1.102) 
&  \\
overdispersion heterogeneity $\sigma_{\varphi}$ 
& 0.804 (0.280, 2.009) 
&  
& 1.214 (0.700, 2.048) 
&  \\
\midrule

\textbf{Mild asthma (16 studies)} & & & & \\
rate intercept $\beta_0$ 
& -0.477 (-2.044, 1.128) 
&  
& -0.501 (-1.717, 0.751) 
&  \\
rate slope $\beta_1$ (publication year) 
& -0.027 (-0.166, 0.114) 
& -2.693 (-15.291, 12.125) 
& -0.028 (-0.136, 0.081) 
& -2.738 (-12.754, 8.410) \\
rate heterogeneity $\sigma_{\lambda}$ 
& 1.752 (1.230, 2.559) 
&  
& 1.385 (1.112, 1.749) 
&  \\
overdispersion $\mu_{\varphi}$ 
& 0.507 (-1.070, 1.521) 
&  
& 0.465 (-0.432, 1.080) 
&  \\
overdispersion heterogeneity $\sigma_{\varphi}$ 
& 1.289 (0.682, 2.390) 
&  
& 1.197 (0.727, 1.969) 
&  \\
\midrule

\textbf{Moderate asthma (32 studies)} & & & & \\
rate intercept $\beta_0$ 
& 0.603 (-0.403, 1.679) 
&  
& 0.617 (-0.425, 1.707) 
&  \\
rate slope $\beta_1$ (publication year) 
& -0.042 (-0.123, 0.037) 
& -4.099 (-11.564, 3.764) 
& -0.042 (-0.125, 0.039) 
& -4.151 (-11.740, 4.009) \\
rate heterogeneity $\sigma_{\lambda}$ 
& 1.332 (1.013, 1.823) 
&  
& 1.385 (1.112, 1.749) 
&  \\
overdispersion $\mu_{\varphi}$ 
& 0.447 (-0.746, 1.237) 
&  
& 0.465 (-0.432, 1.080) 
&  \\
overdispersion heterogeneity $\sigma_{\varphi}$ 
& 1.141 (0.565, 2.098) 
&  
& 1.197 (0.727, 1.969) 
&  \\
\midrule

\textbf{Severe asthma (14 studies)} & & & & \\
rate intercept $\beta_0$ 
& 0.480 (-0.927, 2.628) 
&  
& 1.796 (-0.992, 5.279) 
&  \\
rate slope $\beta_1$ (publication year) 
& -0.016 (-0.143, 0.075) 
& -1.617 (-13.344, 7.811) 
& -0.087 (-0.300, 0.091) 
& -8.339 (-25.903, 9.521) \\
rate heterogeneity $\sigma_{\lambda}$ 
& 0.726 (0.485, 1.195) 
&  
& 1.385 (1.112, 1.749) 
&  \\
overdispersion $\mu_{\varphi}$ 
& -3.650 (-8.933, 0.999) 
&  
& 0.465 (-0.432, 1.080) 
&  \\
overdispersion heterogeneity $\sigma_{\varphi}$ 
& 0.686 (0.032, 2.186) 
&  
& 1.197 (0.727, 1.969) 
&  \\

\bottomrule
\end{tabular}
\end{table}

\clearpage
\subsection{Trends in patient characteristics}\label{sec:patient-characteristic-trends}
 \begin{figure*}[htbp]
	\centering
	\subfloat[age]{\includegraphics[width=.45\linewidth]{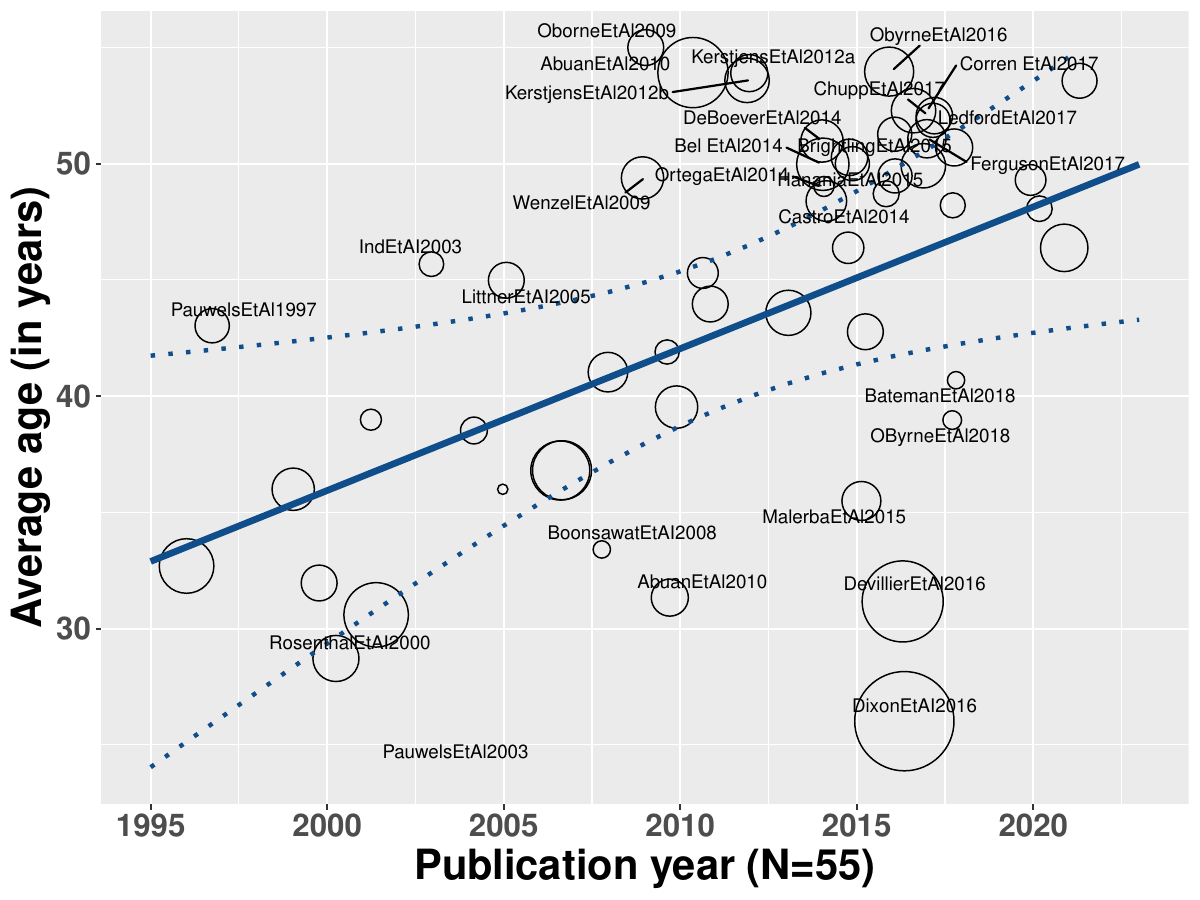}}\hspace{5pt}
	\subfloat[AQLQ score]{\includegraphics[width=.45\linewidth]{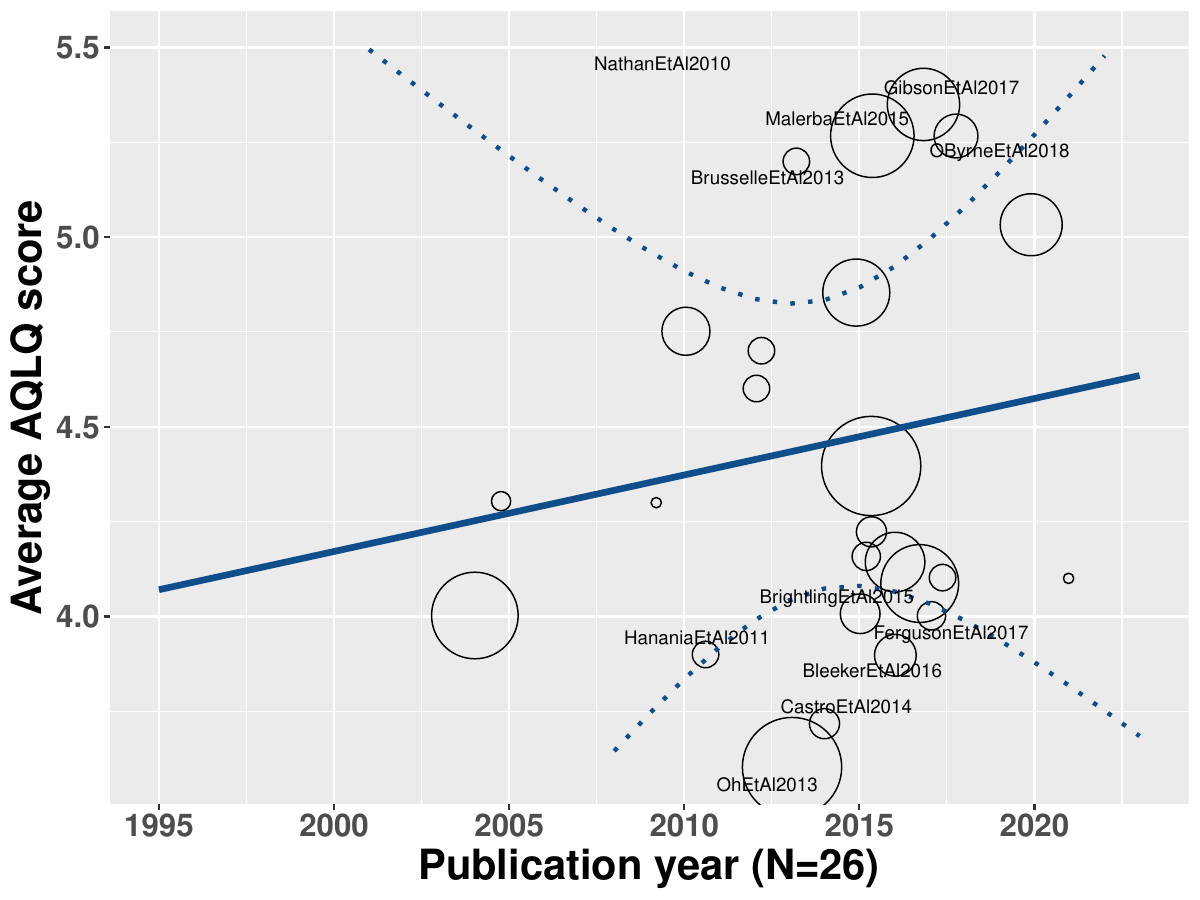}}\\
	\subfloat[{\fevone} (L)]{\includegraphics[width=.45\linewidth]{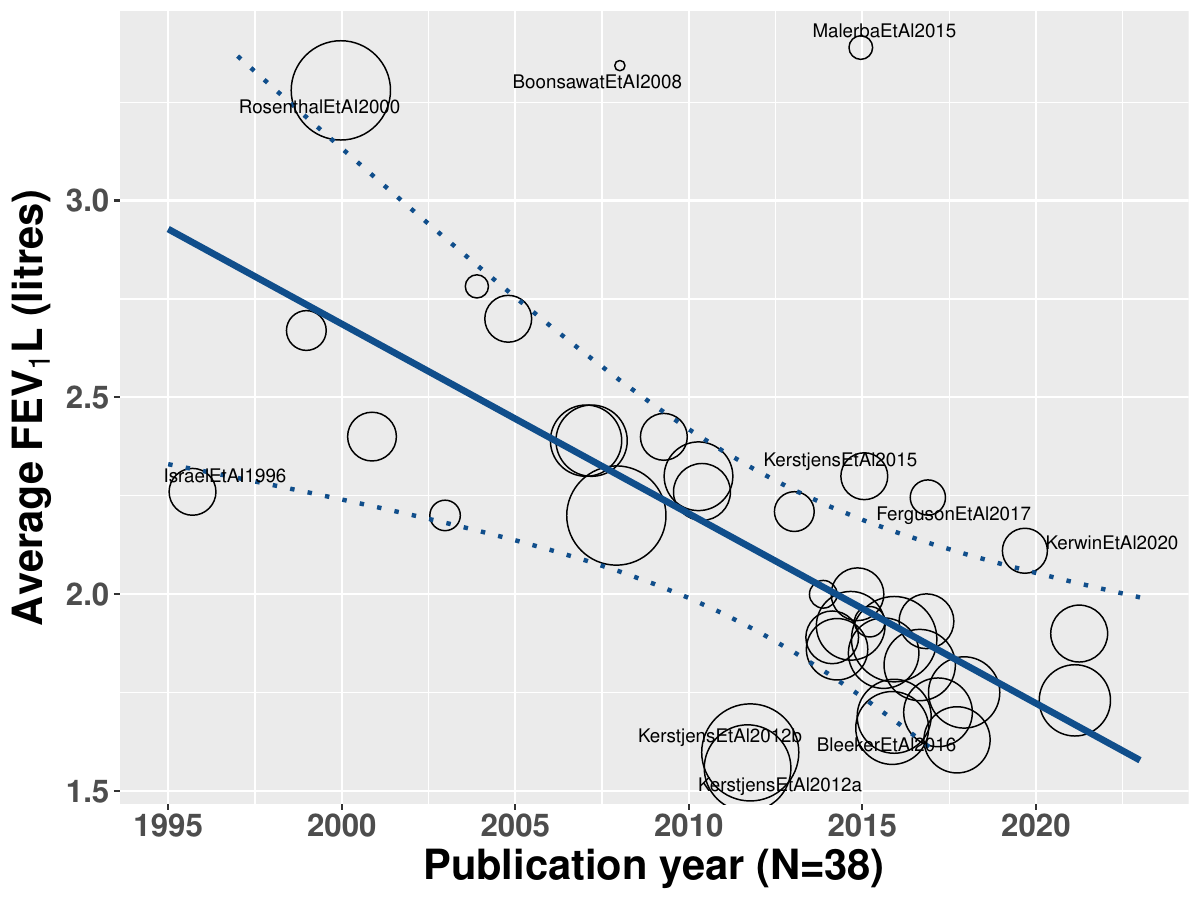}}\hspace{5pt}
	\subfloat[{\fevonepp} (\%)]{\includegraphics[width=.45\linewidth]{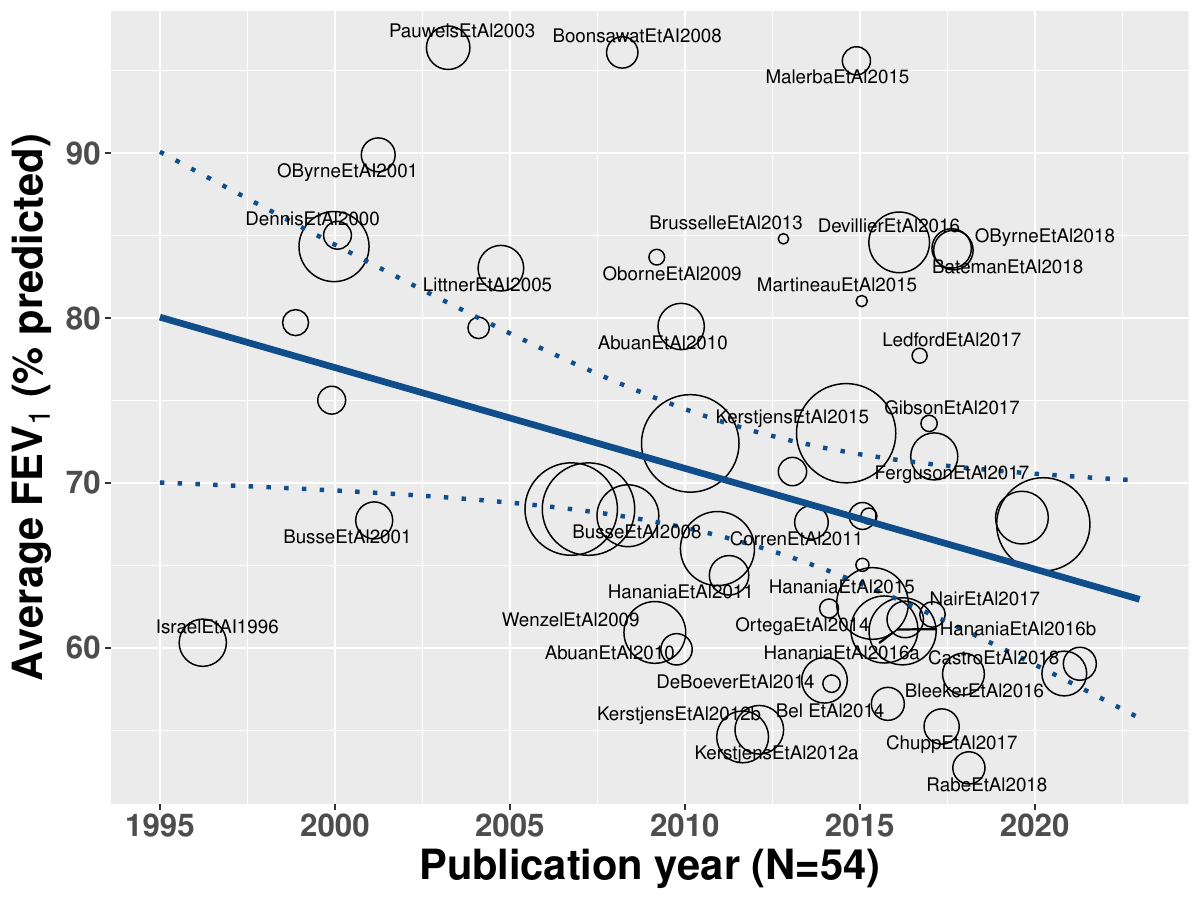}}
	\caption{Time trends in (a) mean age (b) mean AQLQ score (c) mean {\fevone} (L) (d) mean {\fevonepp} (\%) in a mixed-effects meta-regression model with pulication year. Point sizes are proportional to the respective study sizes. Straight lines show predictions 
    as a function of publication year,
    dotted lines the corresponding 95\% CIs.}
    \label{fig:mixed-effect}
\end{figure*}

In order to assess potential confounding by baseline characteristics, we fitted several univariate mixed-effects meta-regression models as illustrated in Figure~\ref{fig:mixed-effect}; the corresponding coefficient estimates are provided in Table~\ref{table:mixed-effect} in Appendix~\ref{AppendixC}. In these models, (a)~mean age, (b)~mean AQLQ~score, (c)~mean {\fevone}, and (d)~mean {\fevonepp} were regressed on publication year (centered at the year~2000). The analyses included (a)~55~studies, (b)~26~studies, (c)~38~studies, and (d)~54~studies, respectively. Estimated effects of publication year varied across outcomes. A positive time trend was observed for mean patient age at baseline ($\hat{\beta}_1=0.610$, 95\% CI: 0.108, 1.113), corresponding to an increase of approximately 6.1 years per decade. By contrast, negative trends were estimated for mean~{\fevone} ($\hat{\beta}_1=-0.048$, 95\% CI: $-0.081$, $-0.015$) and mean~{\fevonepp} ($\hat{\beta}_1=-0.611$, 95\% CI: $-1.173$, $-0.049$).

For lung function measures, the baseline average {\fevone} decreased by approximately 0.48 litres per decade, while {\fevonepp} decreased by approximately 6.1 percentage points per decade. These findings indicate that, over time, studies have tended to include participants with lower baseline lung function. Importantly, these trends may reflect evolving study populations rather than true population-level deterioration in lung function. The observed decline in baseline {\fevone} measures could be attributable to changes in inclusion criteria, an increased focus on moderate-to-severe asthma populations, or shifts in therapeutic standards and background treatment across study periods. Additional regression diagnostics and graphical displays of these associations are provided in Figures~\ref{fig:linear-placebo} and~\ref{fig:linear-severity} in Appendix~\ref{AppendixC}. These exploratory analyses are not intended to establish causal explanations but indicate that temporal changes in study populations could contribute to heterogeneity in exacerbation rates.

\subsection{Implications for sample size determination and trial design}\label{sec:sample-size-implications}
Building on the hierarchical regression framework described above, the estimated
placebo exacerbation rate and overdispersion parameter can be used to inform
sample size planning for future trials. 
For the illustrative sample size calculations, we considered the motivating phase~III \textsc{navigator} trial \cite{Navigator} introduced in Section~\ref{sec:AsthmaExample}.
We considered a clinically relevant rate ratio of 0.5 (i.e., $\lambda_2(t)=0.5\,\lambda_1(t)$), and evaluated the placebo rate at calendar year $t=2017$.
This year was chosen to roughly align with the trial's planning period.

The trial enrolled adults and adolescents with severe, uncontrolled asthma
(only 8\% of patients eventually were adolescents) \citep{MenziesGowEtAl2020}.
Patients were randomized 1:1 to receive \emph{tezepelumab} or placebo as
add-on therapy while continuing their prescribed medium- or high-dose ICS plus
at least one additional controller medication, with or without oral
corticosteroids, over a 52-week treatment period.

We considered three complementary approaches to incorporating uncertainty in the event rate and overdispersion parameters. In all three approaches, the planning quantities were based on the posterior predictive distributions of the control-group event rate and overdispersion parameter for a future trial, including future study-level random effects representing residual between-study heterogeneity.

First, a \emph{point-based design} used the posterior medians of the predicted control-group event rate~$\lambda_1(t)$ and overdispersion parameter~$\varphi$, yielding a plug-in estimate of the required sample size. Second, an \emph{assurance design} averaged the conditional frequentist power over the joint posterior predictive distribution of these planning parameters. The assurance-based sample size was defined as the smallest sample size for which the expected conditional power reached the targeted level.

Third, a \emph{joint posterior design} propagated both posterior parameter
uncertainty and future study-level variability in the event rate and
overdispersion parameter into the sample size calculation. Plugging draws from
their joint posterior predictive distribution into the sample size formula in
Equation~(\ref{eqn:NBSampleSize}) yielded a corresponding distribution of the
required sample size~$n$. The 90th percentile of this distribution was used as
a conservative design specification, corresponding to a sample size that
achieved the targeted conditional power in at least 90\% of the posterior
predictive scenarios.

\begin{figure}[tbp]
    \centering
    \includegraphics[width=0.95\linewidth]{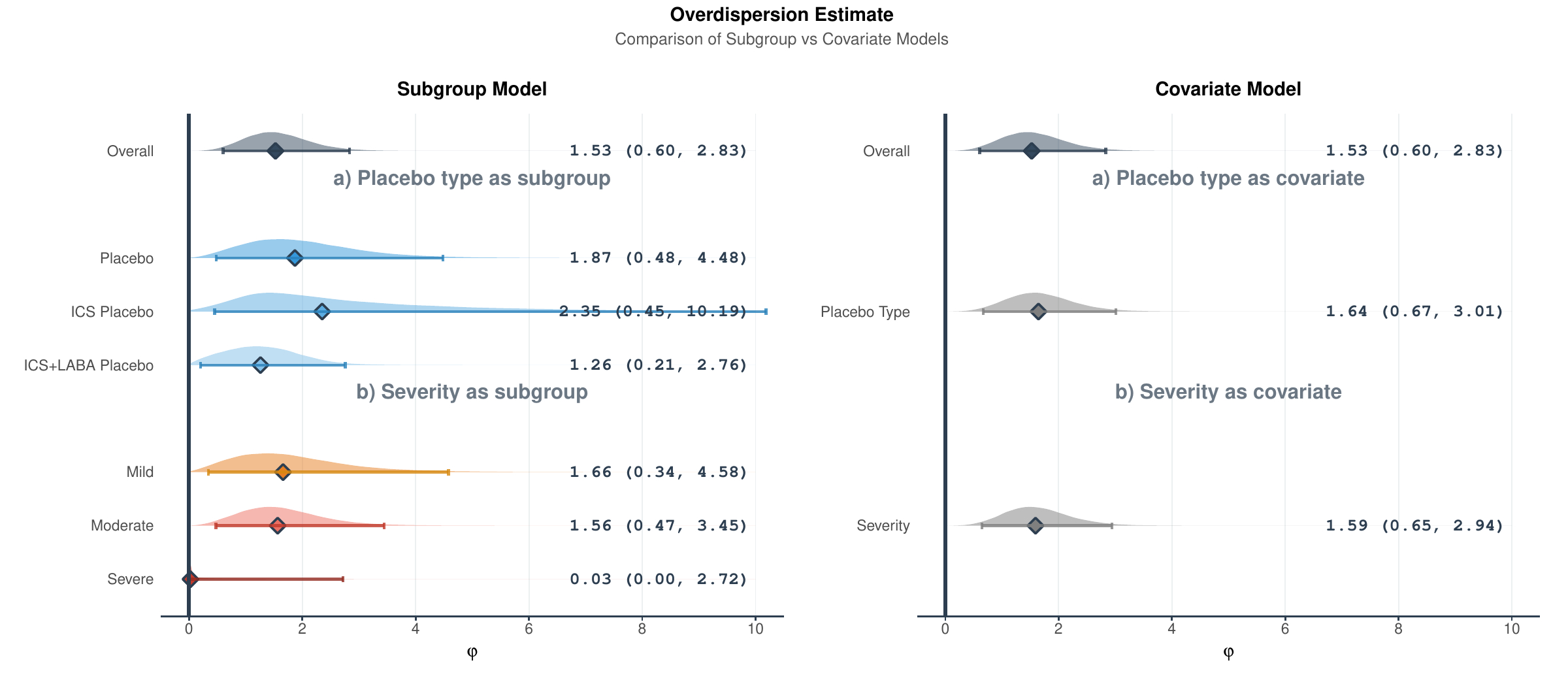}
    \caption{Posterior estimates of the negative binomial overdispersion
    parameter $\varphi$ under subgroup-specific and covariate-based models.
    Estimates are shown for the overall dataset, placebo-type analyses and
    asthma-severity analyses. In the subgroup-specific analyses, $\varphi$ was
    estimated separately within each subgroup, whereas in the covariate-based
    analyses a common hierarchical overdispersion distribution was used within
    each covariate model. Points and horizontal intervals show posterior
    summaries and 95\% credible intervals.}
    \label{fig:phi_forestplot}
\end{figure}

The overdispersion estimates derived from the historical data and used in the sample size calculations are shown in Figure~\ref{fig:phi_forestplot}.
Separate analyses by subgroup only allow to estimate~$\varphi$ with great uncertainty, particularly in smaller subgroups.
In contrast, the joint analyses accounting for subgroups via covariates for the event rate yielded tighter constraints on the common overdispersion estimates. 
This distinction is important for trial design, since greater overdispersion (-uncertainty)
directly impacts on the required sample size under the negative binomial model; 
implementing a common random effect across subgroups them might be worth considering as long as this seems plausible.

\begin{table*}[!htbp]
\centering
\scriptsize
\setlength{\tabcolsep}{3pt}

\caption{Sample size implications under a negative binomial model. The
control-group event rate was evaluated at $t=2017$, with calendar time centered
at $t_0=2000$, assuming a clinically relevant rate ratio of
$\lambda_2(t)/\lambda_1(t)=0.5$. Results are shown for both subgroup-specific
and covariate-based models. The point-based calculations use posterior medians
of the posterior predictive control-group event rate and overdispersion
parameter for a future trial. Assurance-based calculations average conditional
power over their joint posterior predictive distribution, whereas the joint
posterior calculations report the 90th percentile of the resulting sample-size
distribution. Other design parameters were a two-sided significance level of
$\alpha=0.05$, target power of $1-\beta=0.80$, and follow-up duration of
$d=1$ year. Reported sample sizes are per group and were rounded up to the next
integer.}
\label{tab:ss_pred}

\begin{tabular}{llrrrrrrrr}
\hline
Group & Scenario
& \multicolumn{4}{c}{Subgroup model}
& \multicolumn{4}{c}{Covariate model} \\
\cmidrule(lr){3-6}
\cmidrule(lr){7-10}
&
& $\beta_0$ & $\beta_1$ & $\varphi$ & $n$
& $\beta_0$ & $\beta_1$ & $\varphi$ & $n$ \\
\hline

Overall
& Plug-in estimate
& 0.194 & -0.018 & 1.526 & 105
& 0.195 & -0.018 & 1.525 & 104 \\
& Assurance
& -- & -- & -- & 210
& -- & -- & -- & 210 \\
& Joint posterior
& -- & -- & -- & 585
& -- & -- & -- & 586 \\
\hline

True placebo
& Plug-in estimate
& 0.189 & 0.031 & 1.869 & 85
& 0.116 & 0.032 & 1.645 & 79 \\
& Assurance
& -- & -- & -- & 190
& -- & -- & -- & 160 \\
& Joint posterior
& -- & -- & -- & 567
& -- & -- & -- & 419 \\
\hline

ICS-placebo
& Plug-in estimate
& 0.568 & -0.084 & 2.350 & 193
& 0.385 & -0.075 & 1.645 & 173 \\
& Assurance
& -- & -- & -- & 440
& -- & -- & -- & 350 \\
& Joint posterior
& -- & -- & -- & 1397
& -- & -- & -- & 1060 \\
\hline

ICS--LABA-placebo
& Plug-in estimate
& -0.063 & -0.026 & 1.262 & 123
& 0.033 & -0.030 & 1.645 & 133 \\
& Assurance
& -- & -- & -- & 180
& -- & -- & -- & 270 \\
& Joint posterior
& -- & -- & -- & 406
& -- & -- & -- & 758 \\
\hline

Mild asthma
& Plug-in estimate
& -0.477 & -0.027 & 1.660 & 180
& -0.501 & -0.028 & 1.592 & 182 \\
& Assurance
& -- & -- & -- & 480
& -- & -- & -- & 370 \\
& Joint posterior
& -- & -- & -- & 1805
& -- & -- & -- & 1112 \\
\hline

Moderate asthma
& Plug-in estimate
& 0.603 & -0.042 & 1.564 & 106
& 0.617 & -0.042 & 1.592 & 106 \\
& Assurance 
& -- & -- & -- & 210
& -- & -- & -- & 210 \\
& Joint posterior 
& -- & -- & -- & 551
& -- & -- & -- & 576 \\
\hline

Severe asthma
& Plug-in estimate
& 0.480 & -0.016 & 0.026 & 41
& 1.796 & -0.087 & 1.592 & 88 \\
& Assurance 
& -- & -- & -- & 70
& -- & -- & -- & 180 \\
& Joint posterior
& -- & -- & -- & 143
& -- & -- & -- & 470 \\
\hline

\end{tabular}
\end{table*}

\begin{figure}[!htbp]
    \centering
    \includegraphics[width=0.95\linewidth]{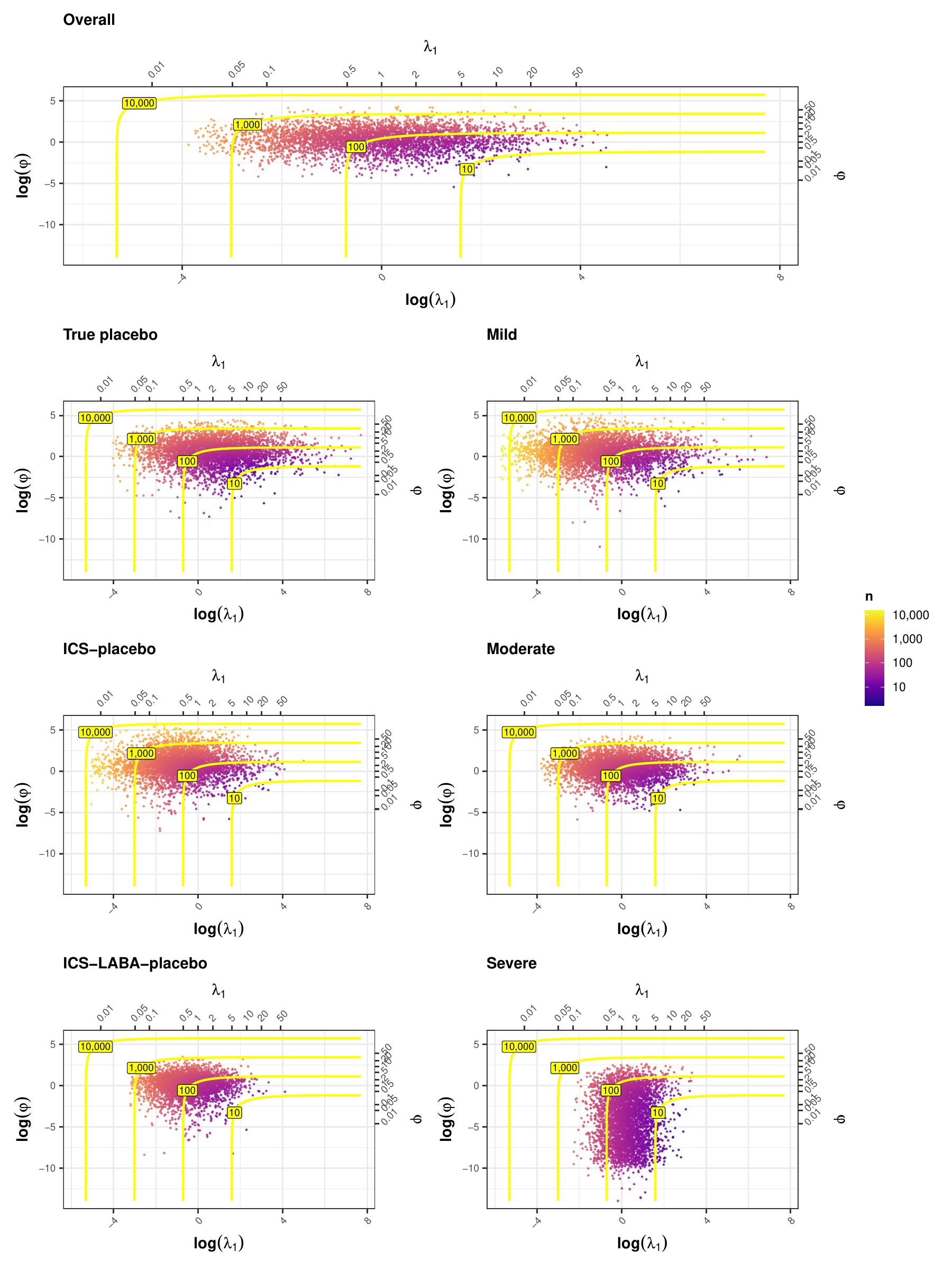}
    \caption{Posterior predictive joint distribution of the placebo exacerbation
    rate $\lambda_1(2017)$ and overdispersion parameter $\varphi$ for a new study
    in 2017 under the subgroup-specific models. Results are shown for the overall
    dataset and for subgroup analyses according to placebo type and asthma severity. Each point represents a posterior predictive draw from the hierarchical model. Colours and contours indicate the corresponding required sample size per group under the assumed rate ratio of 0.5, two-sided $\alpha=0.05$, target power $1-\beta=0.80$, and follow-up duration of $d=1$~year.
    Figure~\ref{fig:sample_size_hist} illustrates the corresponding distributions of the required sample size~$n$.}
    \label{fig:sample_size_2d}
\end{figure}

\begin{figure}[!htbp]
    \centering
    \includegraphics[width=0.95\linewidth]{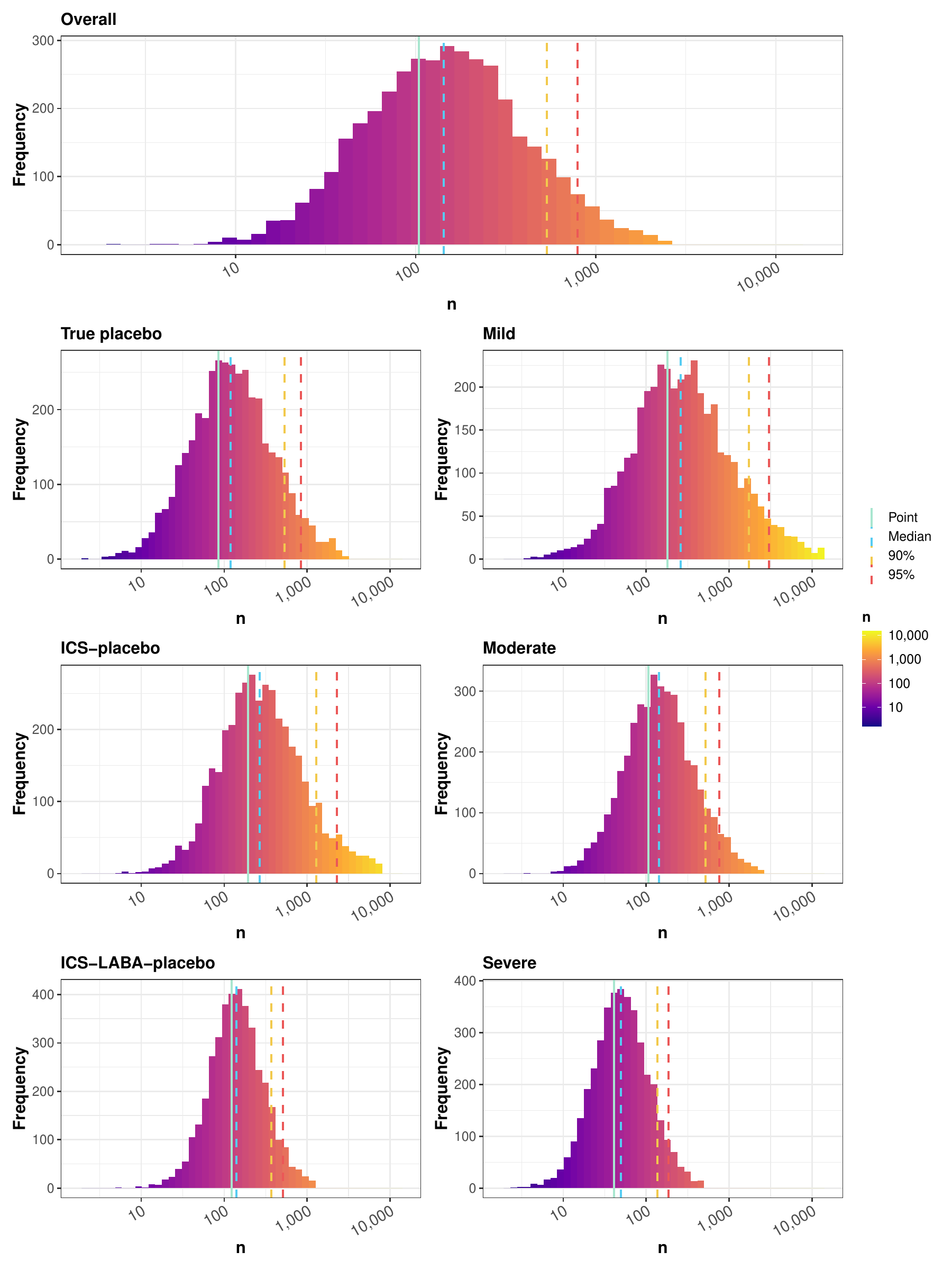}
    \caption{Posterior predictive distributions of the required sample size per group for a new study in 2017 based on the subgroup-specific analyses, corresponding to Figure~\ref{fig:sample_size_2d}.
    Histograms show the distribution of sample sizes obtained by propagating posterior uncertainty in the baseline event rate $\lambda_1$ and overdispersion parameter $\varphi$ under the subgroup-specific negative binomial models. Vertical lines indicate the point estimate based on posterior medians, the posterior median sample size and the 90th percentile of the posterior predictive sample-size distribution. The design parameters are a rate ratio of 0.5, two-sided $\alpha=0.05$, target power $1-\beta=0.80$, and follow-up duration of $d=1$~year.}
    \label{fig:sample_size_hist}
\end{figure}

Figures~\ref{fig:sample_size_2d} and~\ref{fig:sample_size_hist} illustrate
how uncertainty in the baseline exacerbation rate and overdispersion propagates into uncertainty in the required sample size, both in the overall analysis, as well as considering subgroups in separate analyses.
Analogous results from the covariate-based analysis are provided in Appendix~\ref{AppendixD}, where
information is shared across groups through the regression structure. Greater overdispersion and lower event rates both correspond to larger required sample sizes, reflected in the upper tails of the histograms in Figure~\ref{fig:sample_size_hist}.

Table~\ref{tab:ss_pred} lists the sample size estimates based on the overall
and subgroup analyses using the different approaches outlined above. The
point-based calculations produced broadly comparable sample sizes under the
subgroup-specific and covariate-based models for most groups, although a more
pronounced difference was observed for severe asthma. Lower point-based sample
sizes were obtained for groups in which higher event rates might be expected,
particularly the true-placebo and severe-asthma groups. Accounting for posterior 
predictive uncertainty increased the required sample sizes, particularly under 
the joint posterior approach for the ICS-placebo and mild-asthma groups. 
In the overall analysis, the required sample size increased from 104--105 patients 
per group under the point-based design to 210 under the 80\% assurance criterion 
and 585--586 under the joint posterior approach.

For the motivating \textsc{navigator} example, the severe-asthma and
ICS--LABA-placebo results provide the most relevant, although separate,
benchmarks. Under the severe-asthma models, the required sample sizes were
41 and 88 patients per group for the point-based design, 70 and 180 for the
assurance design, and 143 and 470 for the joint posterior design under the
subgroup-specific and covariate-based models, respectively. For the
ICS--LABA-placebo models, the corresponding sample sizes were 123 and 133,
180 and 270, and 406 and 758 patients per group. However, these results should
not be interpreted as estimates for a combined severe-asthma and
ICS--LABA-placebo subgroup, because placebo type and asthma severity were
analysed separately. Moreover, the background therapy permitted in
\textsc{navigator} was broader than the ICS--LABA-placebo category used in
our review.

\section{Discussion}\label{sec:discussion}
The design of clinical trials remains challenging, as it requires specification of certain (``nuisance'') parameters that are essentially unknown before the experiment has actually started \citep{Julious2004}.
Explicit formulation of prior knowledge then is essential for coherent specifications \citep{Lindley1997,BruttiEtAl2014}, and  consultation of external, ``historical'' data then may help predicting reasonable parameter ranges to be expected in a given context \citep{SchmidliNeuenschwanderFriede2017}. Matters get more complicated in case there are relevant external factors to be considered for prediction, such as patient subpopulations or temporal trends \citep{NicholasEtAl2019,AndreasEtAl2022}.

In the present investigation, we consider an example application in asthma, where clinical trials commonly focus on counts of (recurrent) exacerbations as primary endpoints. Since in other disease areas, auch as COPD or multiple sclerosis, event counts in studies' placebo groups have been observed to evolve over time \citep{AndreasEtAl2019,AndreasEtAl2022,SteinvorthEtAl2013,NicholasEtAl2019}, the question arises whether similar trends may be relevant in asthma studies as well. Exacerbation counts in asthma trials are commonly analysed based on negative binomial models \citep{Hilbe2011}, so that for study design information on event rate as well as overdispersion parameters is crucial. To showcase the application of a Bayesian hierarchical meta-regression model to coherently analyze data from past studies' placebo groups, we first conducted a systematic review of published randomized controlled trials in asthma.
Since exacerbation counts are reported in different ways (e.g., sometimes event counts are given, sometimes numbers of patients with and without events are reported), a somewhat complex model specification was required for a joint analysis \citep{RoeverAndreasFriede2016}. In the analysis, we adjusted for (potential) time trends, as well as for additional study-level features, namely, types of placebo groups (based on allowed add-on medications) as well as disease severity.
Technically, the analysis constitutes a meta-regression application with joint prediction of rate and overdispersion.
Different approaches for considering nuisance parameter estimates were followed, from simple plug-in use of point estimates to more sophisticated procedures to fully propagate estimation uncertainty. In addition, different strategies for considering different types of studies were implemented, either restricting the external data to the corresponding subgroup, or applying a coherent model including all studies, while allowing for group-specific rates and a common overdispersion.

Application to the present data set yielded slightly different sample size estimates based on the different assumptions implemented. Taking a broad view and considering the data as a whole yields lower event rate estimates and greater sample sizes. When focusing on severe asthma, exacerbation rates tend to be more frequent. However when restricting analysis to studies of this type alone, overdispersion is estimated as rather low (although with substantial uncertainty), while coherent consideration of the full study set yields a larger (and probably more realistic) overdispersion estimate.
Simple plug-in estimation suggests smaller sample sizes, while more comprehensive estimates based on assurance or a (conservative) 90\% quantile for the sample size both yield greater required patient numbers.

A potential temporal trend was considered in the analysis, suggesting a decline over time, although the direction (increase or decline) was not definite based on the historical data; the annual relative change was estimated as $-1.8\%$ (95\% CrI: $-7.5\%$ to $4.3\%$). Similarly, different event rates were derived depending on whether the focus was on particular placebo types or disease severities. Estimation (and prediction) of overdispersion remained challenging; even when interest is in a specific placebo type or disease severity, borrowing of information on overdispersion across subgroups by using an overarching model allowing for subgroup-specific rate estimates may be worth while.

An important direction for future research is to combine historically informed planning with adaptive designs using blinded sample size re-estimation during an internal pilot phase \citep{FriedeSchmidli2010}.
Updating estimates of event rates and overdispersion using accumulating trial data may improve robustness to misspecified planning assumptions. A further extension could combine prior information from historical studies with blinded interim data, while allowing for potential differences between the historical and current trial populations.
Model selection or model averaging could also be considered to address uncertainty about temporal trends, relevant covariates, and assumptions governing information sharing across subgroups \citep{RoeverFriede2023,RoeverWandelFriede2019}.
In particular, model averaging could propagate uncertainty about model structure into the joint prediction of event rates and overdispersion, and consequently into sample size planning.

\backmatter


\bmhead{Acknowledgements}
We thank Sebastian Straube, Stefan Andreas, Henrik Watz, and Judith Heinz for their contributions to the systematic review, including study screening, data extraction, and data verification.

\section*{Funding}
Open Access funding enabled and organized by Projekt DEAL. This work was funded by the DZHK (German Center for Cardiovascular Research), funding code: 81Z0300108.

\section*{Competing interests}
The authors declare no competing interests.

\section*{Author contribution}
All the other authors have participated in the systematic reviews; they have edited and reviewed the manuscript.

\section*{Data availability}
The study-level data extracted from the included trials and analysed
in this study are provided in the supplementary file.
\texttt{AsthmaData.csv}.

\noindent

\begin{appendices}

\section{Hierarchical modeling of aggregated recurrent event data}
\label{AppendixA}
In the present analysis, we focus on exacerbation outcomes in asthma trials.
Typical reporting formats include endpoints provided as (i) rates together with
standard errors or confidence intervals, (ii) rates or total event counts only,
(iii) numbers or proportions of patients experiencing at least one event, or
(iv) combinations of total event counts and event-free patient
numbers~\cite{RoeverAndreasFriede2016,AndreasEtAl2019}.

In the analyses reported in the main text, the same hierarchical framework was
implemented in two ways to account for placebo type or asthma severity. First,
in the subgroup-specific analyses, the model was fitted independently within
each subgroup, so that the event-rate regression parameters, between-study
heterogeneity, and overdispersion parameters were estimated using only studies
from the corresponding subgroup. This approach provides estimates that are
directly aligned with the target population but may be less stable when the
number of studies in a subgroup is small.

Second, in the covariate-based analyses, all eligible studies were analysed
jointly, and placebo type or asthma severity was incorporated through
study-level covariates in the event-rate model. The overdispersion parameters
were modelled using a common hierarchical distribution across studies. This
approach allows information to be shared across related groups while adjusting
for systematic differences in event rates. The subgroup-specific and
covariate-based analyses were therefore interpreted as complementary: the
former prioritizes direct relevance to the target subgroup, whereas the latter
improves stability under additional model assumptions.

The second and third levels of the hierarchical model are common to all four
reporting formats. For each study~$i=1,\dots,I$, the study-specific event rate
and overdispersion parameter are modelled as

\begin{equation}
\textbf{Level 2:} \quad
\log(\lambda_i)\mid\boldsymbol{\beta},\sigma_\lambda^2
\sim
\normaldistn(\mathbf{x}_i^\top\boldsymbol{\beta},\sigma_\lambda^2),
\end{equation}
\begin{equation}
\qquad\quad
\log(\varphi_i)\mid\mu_\varphi,\sigma_\varphi^2
\sim
\normaldistn(\mu_\varphi,\sigma_\varphi^2).
\end{equation}

The corresponding prior distributions are

\begin{equation}
\textbf{Level 3:} \quad
\boldsymbol{\beta}\sim\pi(\boldsymbol{\beta}),
\qquad
\sigma_\lambda^2\sim\pi(\sigma_\lambda^2),
\end{equation}
\begin{equation}
\qquad\quad
\mu_\varphi\sim\pi(\mu_\varphi),
\qquad
\sigma_\varphi^2\sim\pi(\sigma_\varphi^2).
\end{equation}

Level 2 specifies the between-study variability through a Bayesian regression
structure relating the study-specific event rate to study-level
covariates~$\mathbf{x}_i$, as well as variability in the overdispersion
parameter. Level 3 specifies prior distributions for the model parameters. The
Level 1 likelihood depends on the reporting format and is specified below.

\paragraph{Case I: Reported recurrent rates with standard errors}

For studies reporting recurrent rates together with associated standard errors
or confidence intervals, the observed outcome can be treated as an estimate of
the underlying event rate. Let~$Rate_i$ denote the reported recurrent rate in
study~$i$, and let~$\sigma_{rate_i}$ denote the reported standard error
of~$\log(Rate_i)$, assumed to be known. The corresponding likelihood is

\begin{equation}
\textbf{Level 1:} \quad
\log(Rate_i)\mid\lambda_i
\sim
\normaldistn\big(\log(\lambda_i),\sigma_{rate_i}^2\big).
\end{equation}

This likelihood provides direct information about the study-specific event
rate~$\lambda_i$, but does not directly inform the study-specific
overdispersion parameter~$\varphi_i$.

\paragraph{Case II: Total recurrent counts only}

For studies reporting only the total number of recurrent events, the observed
outcome can be interpreted as arising from the aggregation of individual
patient event counts. Let~$T_i$ denote the total number of observed recurrent
events in study~$i$, with~$n_i$ patients followed over exposure
duration~$\delta_i$. Let~$\lambda_i$ denote the underlying recurrent event rate
per patient per unit time in study~$i$, and let~$\varphi_i$ denote the
corresponding overdispersion parameter capturing extra-Poisson variability
across patients. The corresponding likelihood is

\begin{equation}
\textbf{Level 1:} \quad
T_i\mid\lambda_i,\varphi_i
\sim
\normaldistn(\mu_{T_i},\sigma_{T_i}^2),
\end{equation}
where
\begin{equation}
\mu_{T_i}=n_i\delta_i\lambda_i,
\qquad
\sigma_{T_i}^2
=
n_i\delta_i\lambda_i
\left(1+\varphi_i\delta_i\lambda_i\right).
\end{equation}

Here, $\mu_{T_i}=n_i\delta_i\lambda_i$ denotes the expected total number of
events obtained by summing individual patient event counts assuming a common
event rate within study~$i$, and
$\sigma_{T_i}^2=n_i\delta_i\lambda_i
\left(1+\varphi_i\delta_i\lambda_i\right)$ represents the corresponding
variance implied by the negative binomial event-count model, allowing for
overdispersion beyond the Poisson assumption.

\paragraph{Case III: Patients without recurrent events}

For studies reporting the number or proportion of patients experiencing no
recurrent events, the observed outcome reflects the probability of remaining
event-free during the follow-up period. Let~$Z_i$ denote the number of patients
without recurrent events in study~$i$, among~$n_i$ patients followed over
exposure duration~$\delta_i$.

Assuming individual patient event counts follow a negative binomial
distribution with event rate~$\lambda_i$ and overdispersion
parameter~$\varphi_i$, the probability of observing zero events for a patient
is given by

\begin{equation}
\pi_{0i}
=
P(X=0)
=
\left(1+\varphi_i\delta_i\lambda_i\right)^{-1/\varphi_i}.
\end{equation}

The corresponding likelihood is

\begin{equation}
\textbf{Level 1:} \quad
Z_i\mid\lambda_i,\varphi_i
\sim
\text{Binomial}(n_i,\pi_{0i}).
\end{equation}

This likelihood is based on the probability of remaining event-free derived
from the negative binomial event-count model.

\paragraph{Case IV: Joint reporting of total recurrent events and event-free
patients}

For studies reporting both the total number of recurrent events and the number
of patients without events, the observed outcomes provide complementary
information about the underlying event-count process. Let~$T_i$ denote the
total number of recurrent events observed in study~$i$, and let~$Z_i$ denote
the number of patients without recurrent events among~$n_i$ patients followed
over exposure duration~$\delta_i$. Let~$\lambda_i$ denote the underlying event
rate per patient per unit time in study~$i$, and let~$\varphi_i$ denote the
overdispersion parameter.

Under the negative binomial event-count model, the probability of observing
zero events for a patient is

\begin{equation}
\pi_{0i}=P(X=0)=
\begin{cases}
\left(1+\varphi_i\delta_i\lambda_i\right)^{-1/\varphi_i},
& \varphi_i>0 \quad \text{(negative binomial case)}, \\[6pt]
\exp(-\delta_i\lambda_i),
& \varphi_i=0 \quad \text{(Poisson case)}.
\end{cases}
\end{equation}

Consequently,

\begin{equation}
P(X>0)=1-\pi_{0i},
\qquad
\theta_i=\expect[X\mid X>0]
=
\begin{cases}
\dfrac{\delta_i\lambda_i}
{1-\left(1+\varphi_i\delta_i\lambda_i\right)^{-1/\varphi_i}},
& \varphi_i>0, \\[10pt]
\dfrac{\delta_i\lambda_i}
{1-\exp(-\delta_i\lambda_i)},
& \varphi_i=0.
\end{cases}
\end{equation}

The corresponding conditional variance is
$\sigma_{X_i}^2=\operatorname{Var}(X\mid X>0)$.
Let~$\mu_i=\delta_i\lambda_i$. Since

\begin{equation}
\expect[X]=\mu_i,
\qquad
\operatorname{Var}(X)=\mu_i(1+\varphi_i\mu_i),
\end{equation}
we have
\begin{equation}
\expect[X^2]
=
\mu_i+(1+\varphi_i)\mu_i^2.
\end{equation}
Because the zero outcome contributes neither to the first nor to the second raw
moment,

\begin{equation}
\expect[X^2\mid X>0]
=
\frac{\mu_i+(1+\varphi_i)\mu_i^2}
{1-\pi_{0i}}.
\end{equation}

Hence,

\begin{equation}
\begin{aligned}
\sigma_{X_i}^2
&=
\expect[X^2\mid X>0]
-
\{\expect[X\mid X>0]\}^2\\
&=
\frac{\mu_i}{1-\pi_{0i}}
+
\frac{
\mu_i^2
\left\{
\varphi_i-(1+\varphi_i)\pi_{0i}
\right\}
}{
(1-\pi_{0i})^2
}.
\end{aligned}
\end{equation}

Equivalently,

\begin{equation}
\sigma_{X_i}^2=
\begin{cases}
\displaystyle
\frac{\delta_i\lambda_i}
{1-(1+\varphi_i\delta_i\lambda_i)^{-1/\varphi_i}}
+
\frac{
(\delta_i\lambda_i)^2
\left\{
\varphi_i
-
(1+\varphi_i)
(1+\varphi_i\delta_i\lambda_i)^{-1/\varphi_i}
\right\}
}{
\left\{
1-(1+\varphi_i\delta_i\lambda_i)^{-1/\varphi_i}
\right\}^2
},
& \varphi_i>0,
\\[12pt]
\displaystyle
\frac{\delta_i\lambda_i}
{1-\exp(-\delta_i\lambda_i)}
-
\frac{
(\delta_i\lambda_i)^2
\exp(-\delta_i\lambda_i)
}{
\left\{
1-\exp(-\delta_i\lambda_i)
\right\}^2
},
& \varphi_i=0.
\end{cases}
\end{equation}

The corresponding joint likelihood is specified by

\begin{equation}
\textbf{Level 1:} \quad
Z_i\mid\lambda_i,\varphi_i
\sim
\text{Binomial}(n_i,\pi_{0i}),
\end{equation}
and
\begin{equation}
\textbf{Level 1 (conditional):} \quad
T_i\mid Z_i,\lambda_i,\varphi_i
\sim
\normaldistn\big(
(n_i-Z_i)\theta_i,\,
(n_i-Z_i)\sigma_{X_i}^2
\big).
\end{equation}

The joint likelihood is constructed via the factorization

\begin{equation}
p(T_i,Z_i\mid\lambda_i,\varphi_i)
=
p(Z_i\mid\lambda_i,\varphi_i)\,
p(T_i\mid Z_i,\lambda_i,\varphi_i),
\end{equation}
with
\begin{equation}
p(Z_i\mid\lambda_i,\varphi_i)
=
\binom{n_i}{Z_i}
\pi_{0i}^{Z_i}
(1-\pi_{0i})^{n_i-Z_i}
\end{equation}
and
\begin{equation}
p(T_i\mid Z_i,\lambda_i,\varphi_i)
=
\frac{1}
{\sqrt{2\pi(n_i-Z_i)\sigma_{X_i}^2}}
\exp\left\{
-\frac{
\left(T_i-(n_i-Z_i)\theta_i\right)^2
}{
2(n_i-Z_i)\sigma_{X_i}^2
}
\right\}.
\end{equation}

Combining the two components yields the approximate joint density

\begin{equation}
\begin{aligned}
p(T_i,Z_i\mid\lambda_i,\varphi_i)
={}&
\binom{n_i}{Z_i}
\pi_{0i}^{Z_i}
(1-\pi_{0i})^{n_i-Z_i}\\
&\times
\frac{1}
{\sqrt{2\pi(n_i-Z_i)\sigma_{X_i}^2}}
\exp\left\{
-\frac{
\left(T_i-(n_i-Z_i)\theta_i\right)^2
}{
2(n_i-Z_i)\sigma_{X_i}^2
}
\right\}.
\end{aligned}
\end{equation}

This formulation allows the model to jointly utilize information from both
total event counts and the number of event-free patients while remaining
consistent with the underlying negative binomial data-generating mechanism.

\newpage
\section{Prisma flow chart}\label{sec:AppendixB}

\renewcommand{\thetable}{B\arabic{table}}
\setcounter{table}{0}

\renewcommand{\thefigure}{B\arabic{figure}}
\setcounter{figure}{0}

\begin{figure}[H]
    \centering
    \includegraphics[width=1\linewidth]{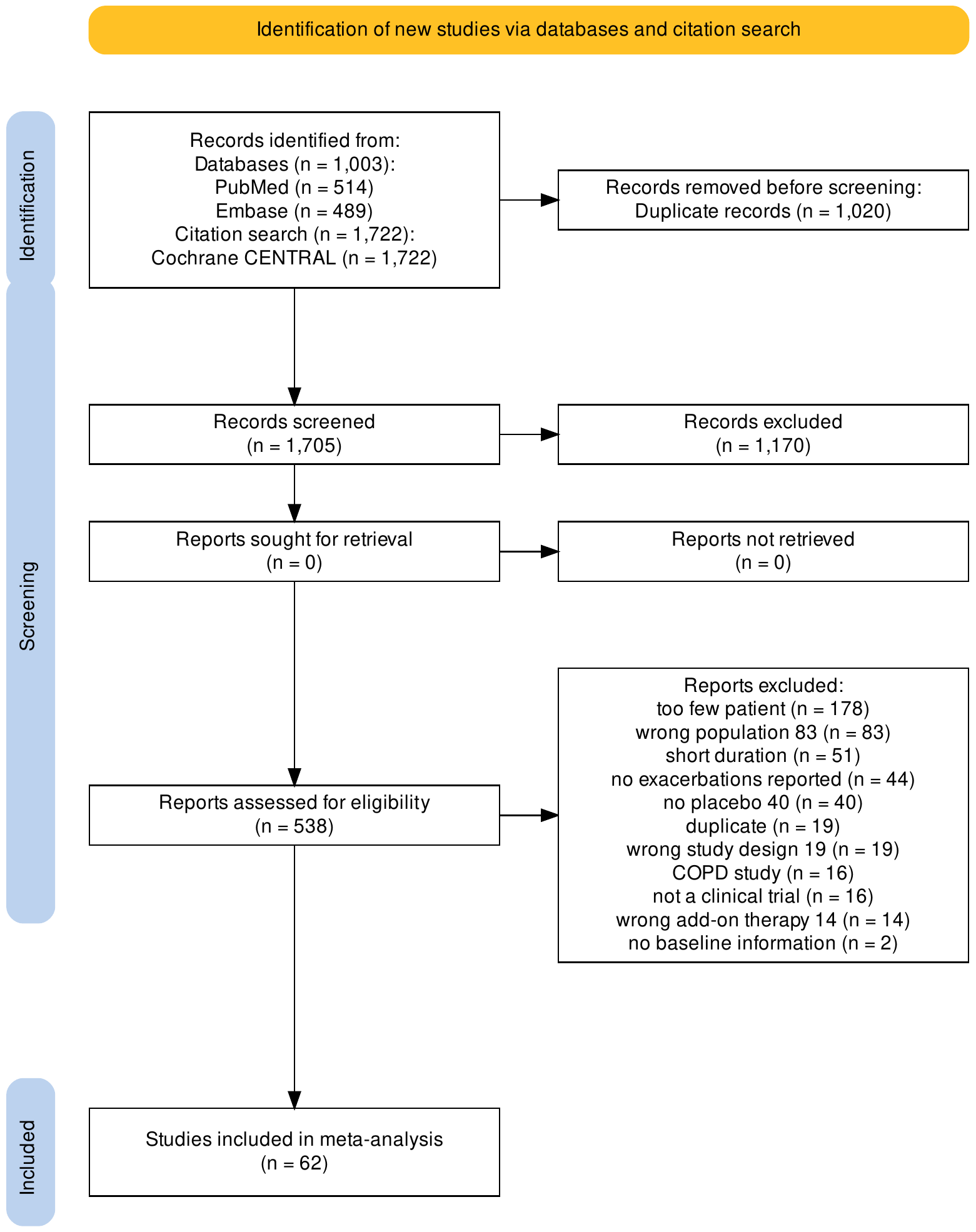}
    \caption{Systematic review: One record contains multiple studies, here we contained 58 records and 62 studies}
    \label{fig:PrismaFlowChart}
\end{figure}

\section{Temporal changes }
\label{AppendixC}

\renewcommand{\thetable}{C\arabic{table}}
\setcounter{table}{0}

\renewcommand{\thefigure}{C\arabic{figure}}
\setcounter{figure}{0}

\begin{table}[htbp]
\centering
\caption{Coefficient estimates, corresponding 95\% CIs, and $p$-values for the associations of publication year with (a) mean age, (b) mean AQLQ score, (c) mean {\fevone} (L), and (d) mean {\fevonepp} (\%) in mixed-effects meta-regression models.}
\begin{tabularx}{\linewidth}{lXr}
\hline
Parameter & Estimate (95\% CI) & $p$-value \\ \hline
Age (55 studies)   & 0.610 (0.108, 1.113)     & 0.017*  \\
AQLQ (26 studies)  & 0.020 (-0.076, 0.116)    & 0.681   \\
{\fevone} (L) (38 studies) & -0.048 (-0.081, -0.015) & 0.004**  \\
{\fevonepp} (\%) (54 studies) & -0.611 (-1.173, -0.049) & 0.033*  \\ \hline
\end{tabularx}

\footnotesize
\textit{Note:} * indicates statistical significance at the 0.05 level; ** at the 0.01 level.

\label{table:mixed-effect}
\end{table}

 \begin{figure}[H]
	\centering
	\subfloat[AGE]{\includegraphics[width=.45\linewidth]{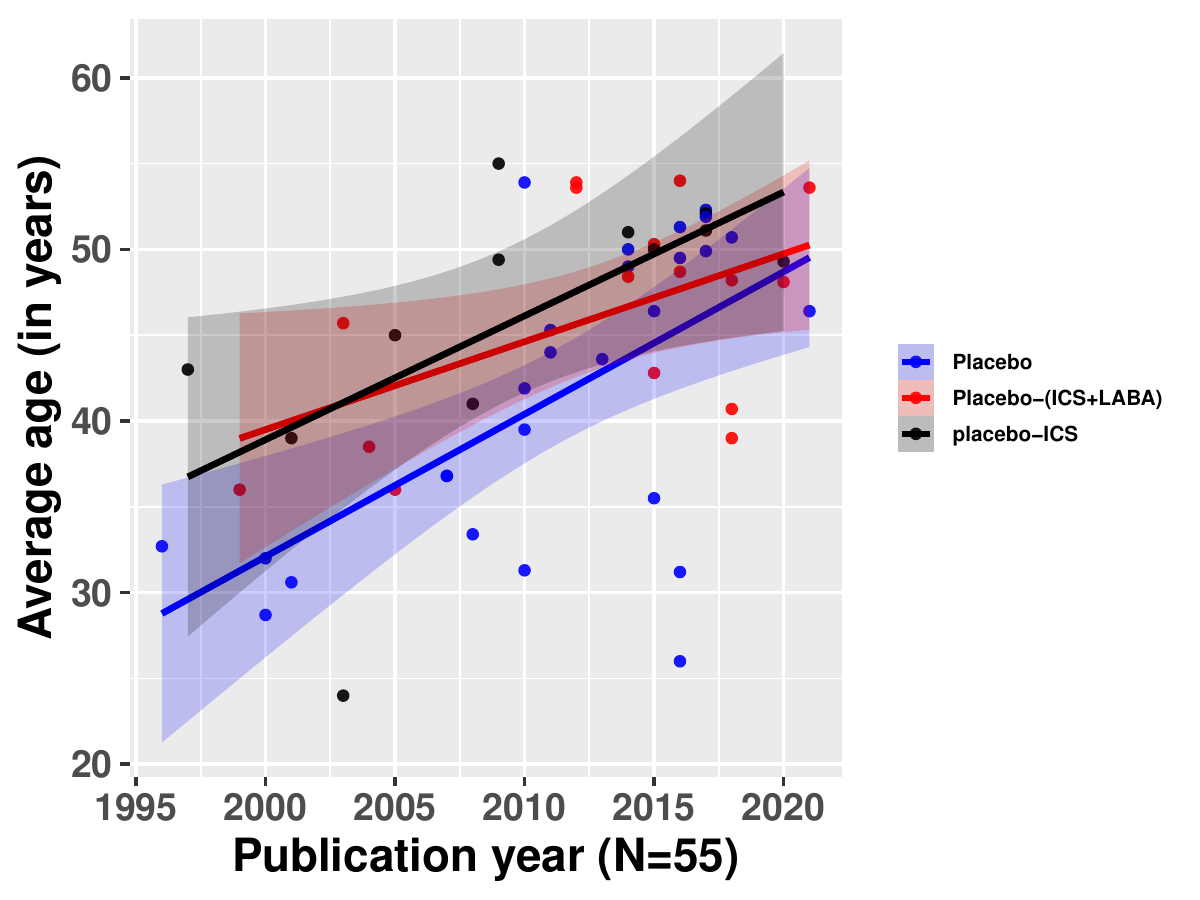}}\hspace{5pt}
	\subfloat[AQLQ]{\includegraphics[width=.45\linewidth]{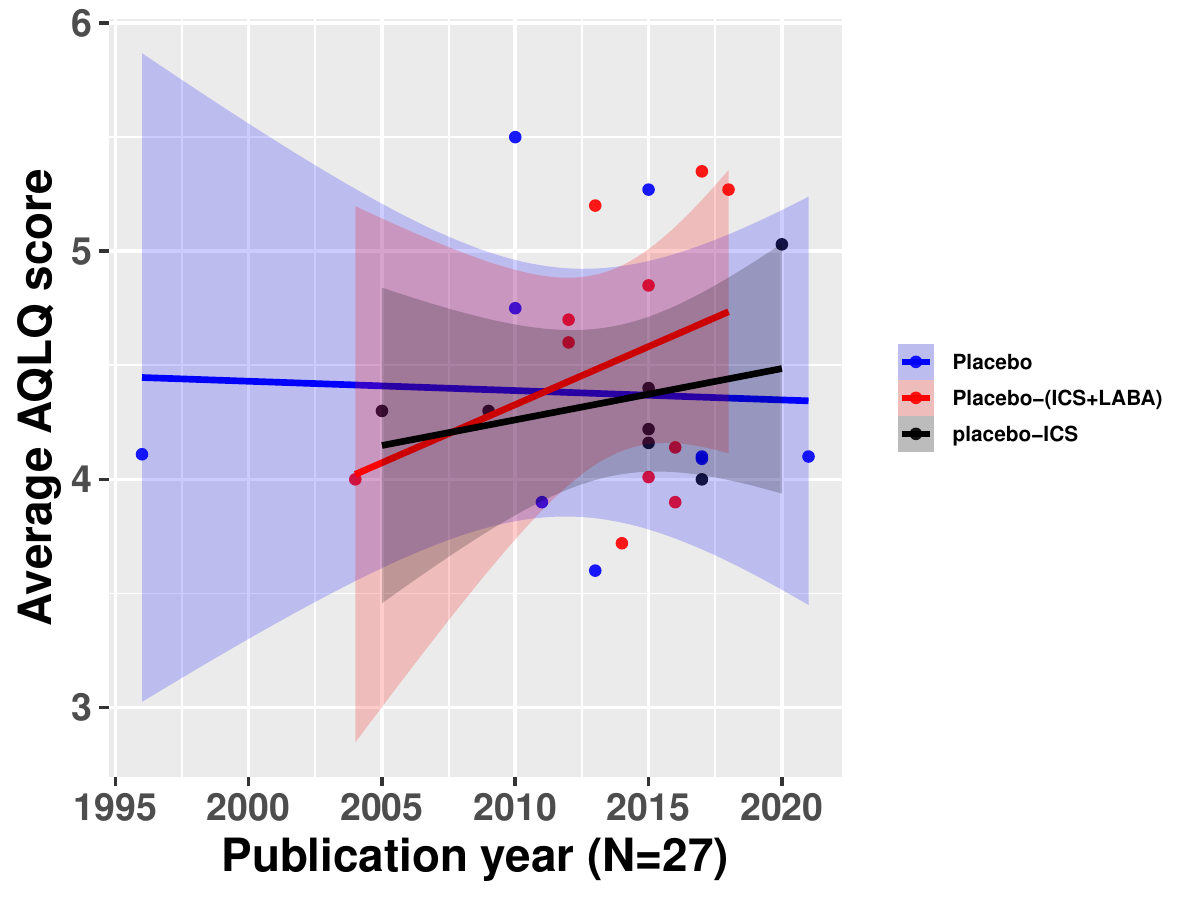}}\\
	\subfloat[{\fevone} (L)]{\includegraphics[width=.45\linewidth]{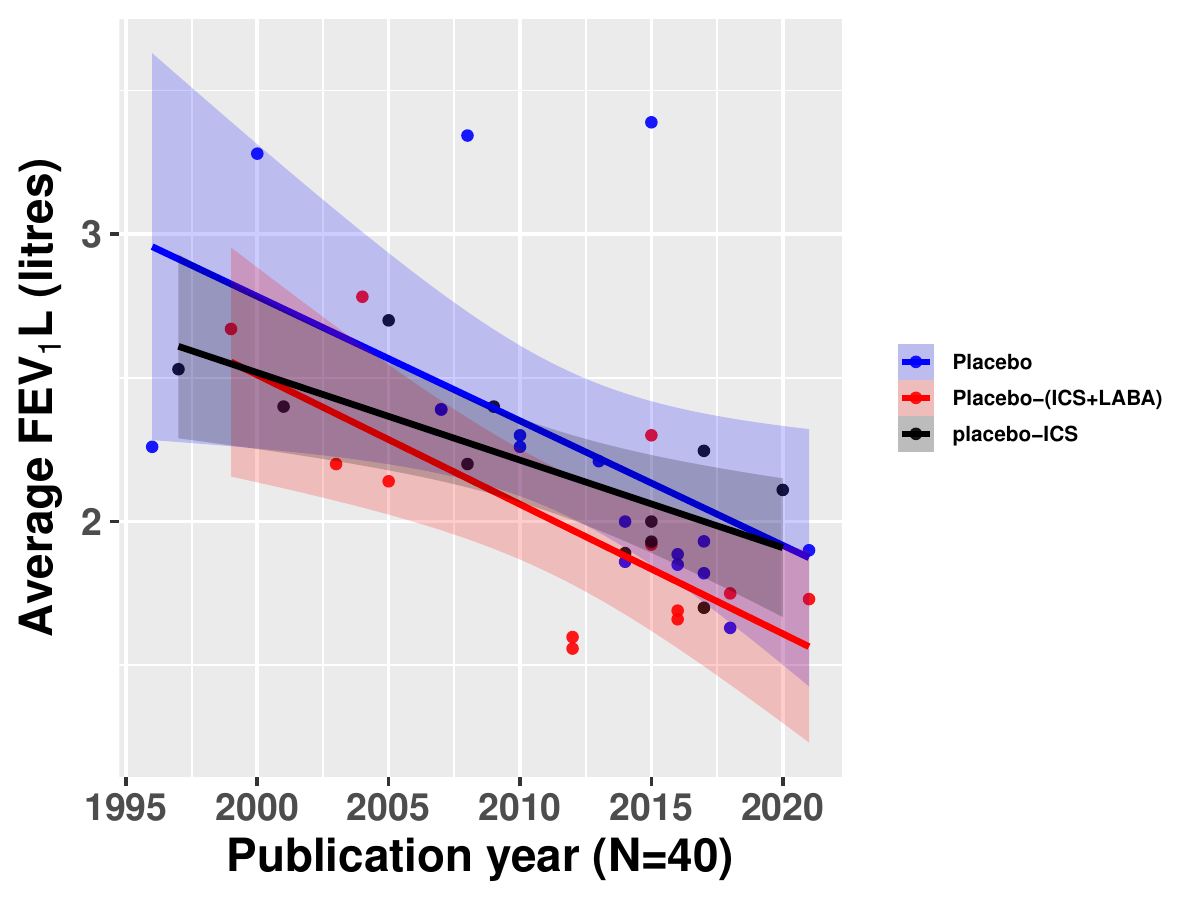}}\hspace{5pt}
	\subfloat[{\fevonepp} (\%)]{\includegraphics[width=.45\linewidth]{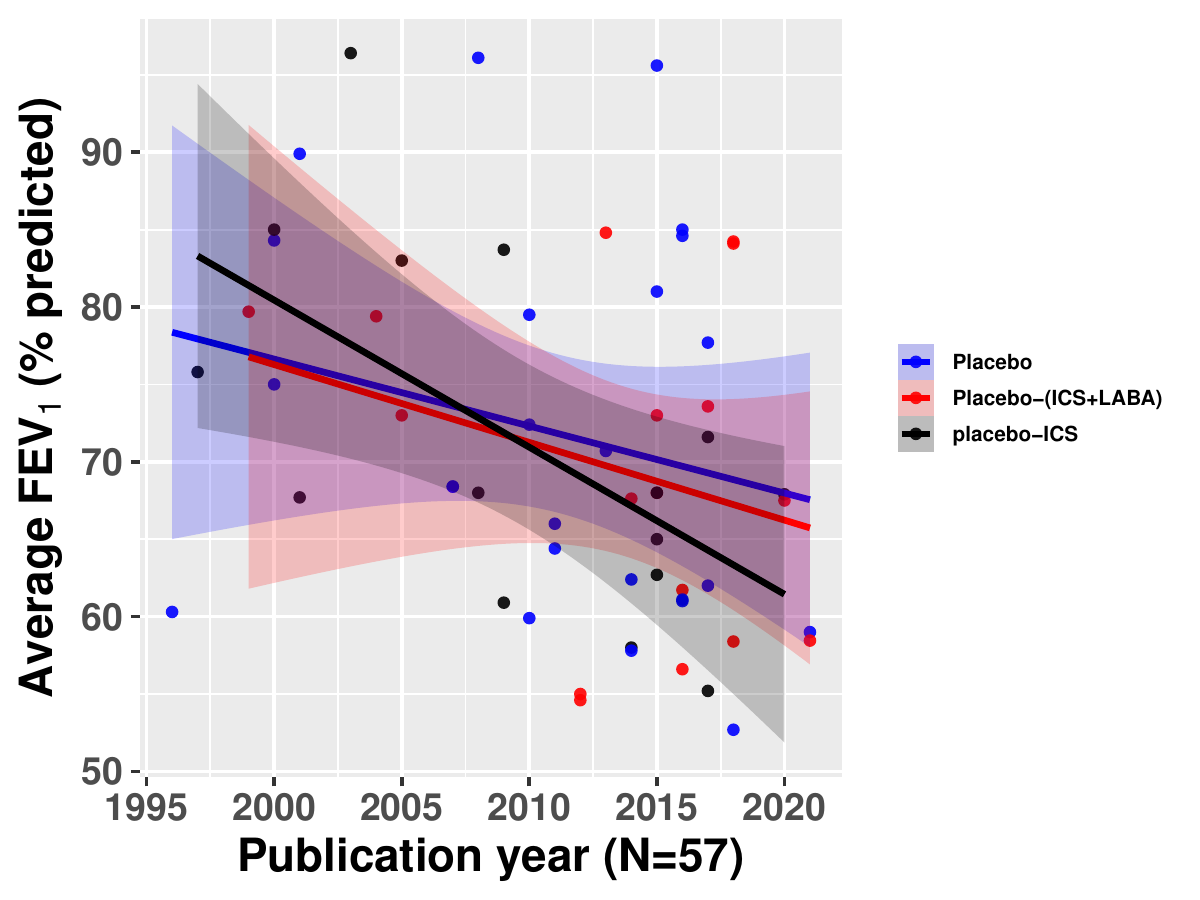}}
	\caption{Trends of publication year on the (a) Average age (b) Average Asthma Quality of Life Questionnaire (c) Average Forced Expiratory Volume in one second as a percentage of Predicted (d) Average Forced Expiratory Volume in one second (litres) in a linear-regression model with these two moderators stratified by placebo group. Straight lines represent the models' predictions at the median level of the respective other explanatory variable, while gray-shaded regions depict the corresponding 95\% CIs.}
    \label{fig:linear-placebo}
\end{figure}

 \begin{figure}[H]
	\centering
	\subfloat[AGE]{\includegraphics[width=.45\linewidth]{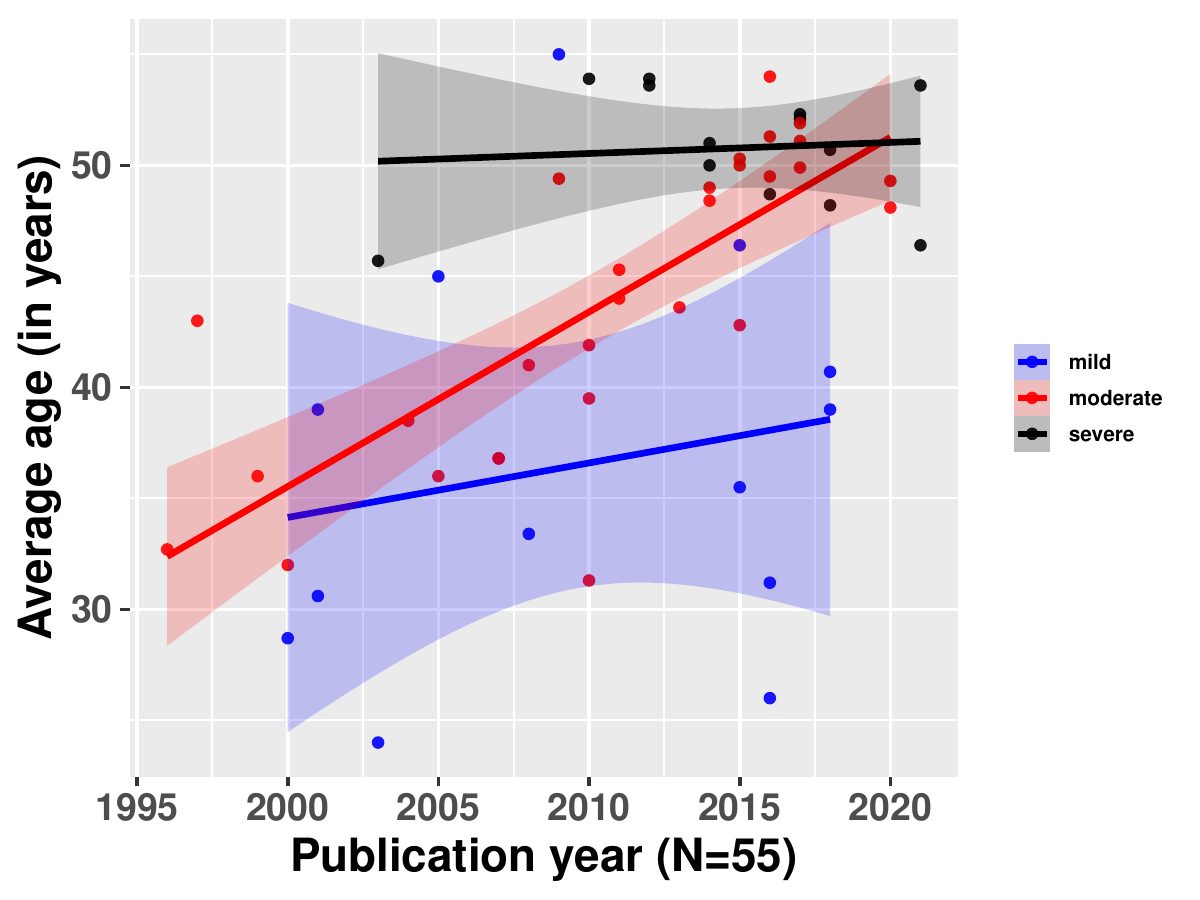}}\hspace{5pt}
	\subfloat[AQLQ]{\includegraphics[width=.45\linewidth]{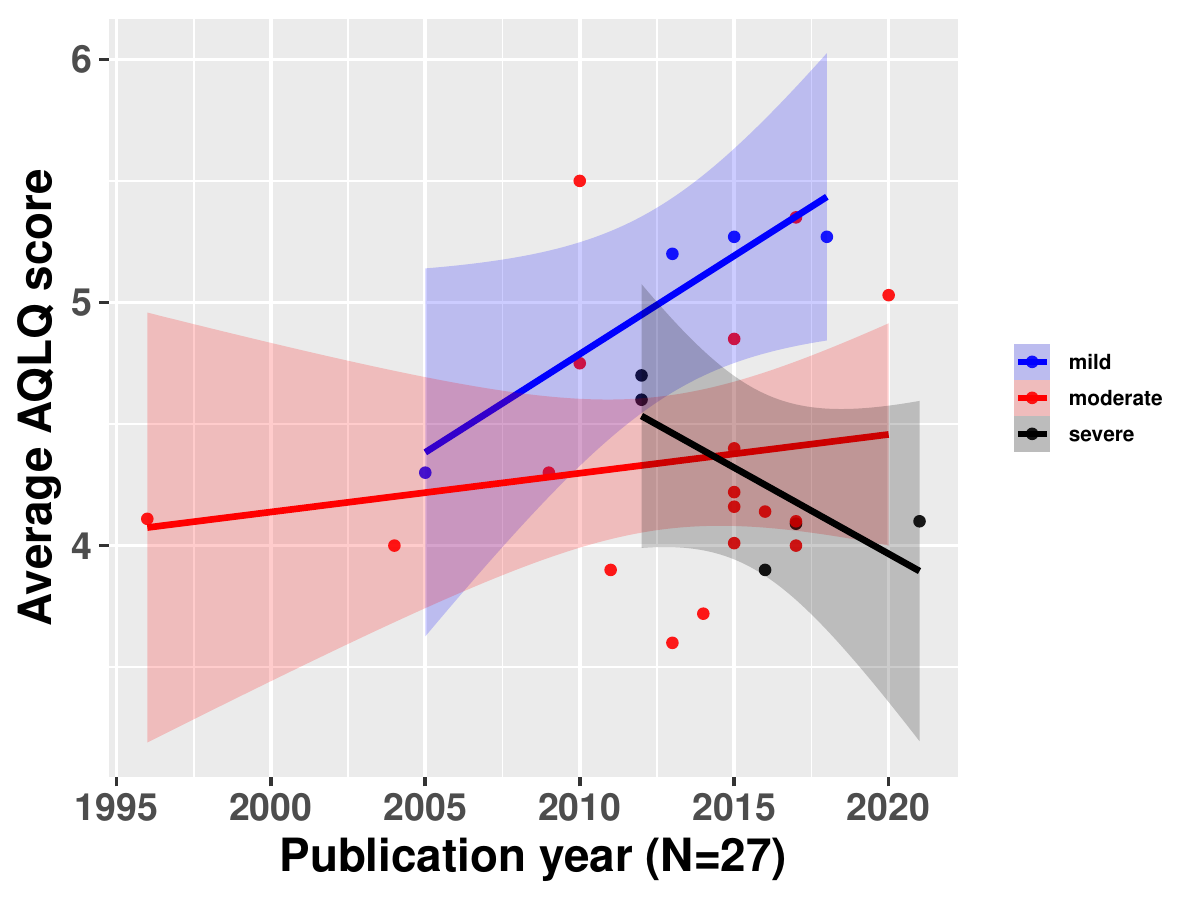}}\\
	\subfloat[{\fevone} (L)]{\includegraphics[width=.45\linewidth]{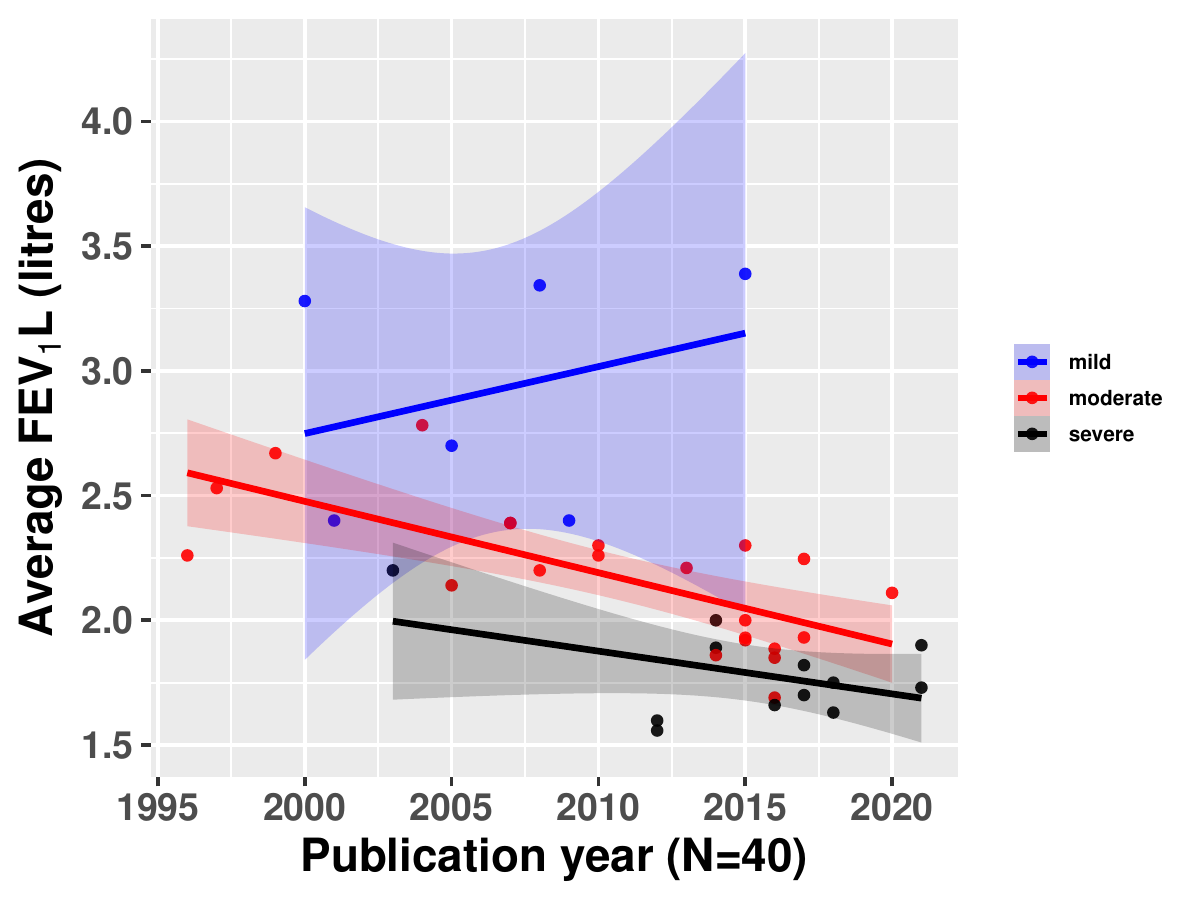}}\hspace{5pt}
	\subfloat[{\fevonepp} (\%)]{\includegraphics[width=.45\linewidth]{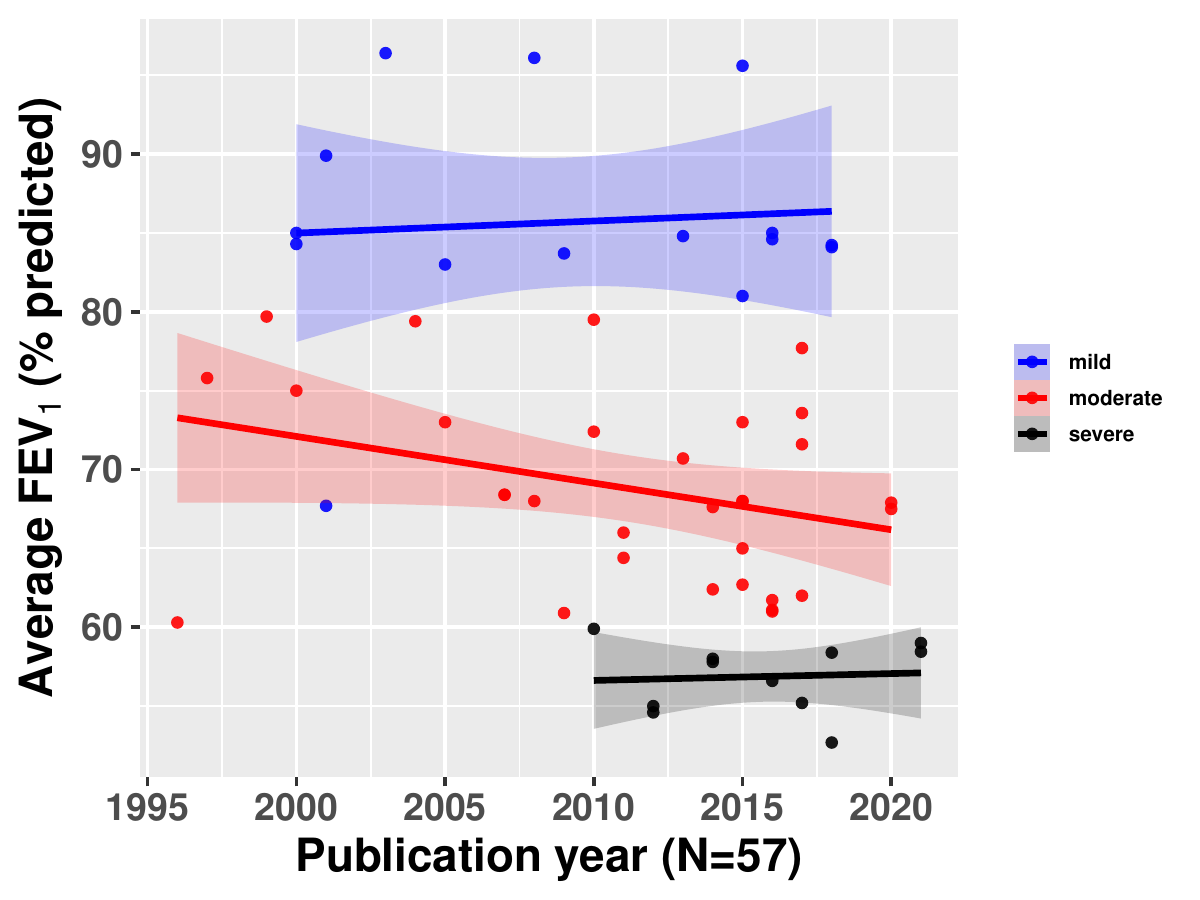}}
	\caption{Trends of publication year on the (a) Average age (b) Average Asthma Quality of Life Questionnaire (c) Average Forced Expiratory Volume in one second (litres) (d) Average Forced Expiratory Volume in one second as a percentage of Predicted in a linear-regression model with these two moderators stratified by asthma severity. Straight lines represent the models' predictions at the median level of the respective other explanatory variable, while gray-shaded regions depict the corresponding 95\% CIs.}
    \label{fig:linear-severity}
\end{figure}

In addition to the mixed-effects meta-regression models, we conducted exploratory linear regression analyses stratified by placebo type and asthma severity in order to visually assess potential subgroup-specific temporal patterns. These models regressed baseline study characteristics on publication year within each subgroup and are presented in the corresponding stratified figures. When stratified by placebo type in Figure~\ref{fig:linear-placebo}, temporal patterns differed across clinical characteristics but remained broadly consistent in direction. In all placebo strata, baseline age showed an increasing trend over publication years, indicating that more recent trials tended to enroll older participants. For lung function measures, both {\fevone} and {\fevonepp} generally declined over time, with the most pronounced decreases observed in ICS-placebo studies, while trends were more modest in true placebo and ICS-LABA-placebo trials. In contrast, AQLQ scores did not demonstrate a clear temporal trend across placebo groups. Although the magnitude of slopes varied between strata, these findings suggest gradual shifts in baseline population characteristics over time rather than abrupt differences attributable solely to placebo classification.

Stratification by asthma severity in Figure~\ref{fig:linear-severity} revealed similarly heterogeneous but interpretable temporal patterns. Baseline age increased over time across mild, moderate, and severe asthma strata. Lung function measures showed downward trends across severity categories, with the decline appearing most pronounced in moderate asthma and less marked in severe asthma. AQLQ scores again showed limited evidence of systematic temporal change. While confidence bands around several estimates were wide, the consistent direction of age increase and lung function decline across strata suggests that trial populations may have evolved over time, potentially reflecting changes in recruitment practices, eligibility criteria, or underlying disease profiles within severity categories.

\section{Sample Size }\label{AppendixD}

\renewcommand{\thetable}{D\arabic{table}}
\setcounter{table}{0}

\renewcommand{\thefigure}{D\arabic{figure}}
\setcounter{figure}{0}

 \begin{figure}[H]
    \centering
    \includegraphics[width=1\linewidth]{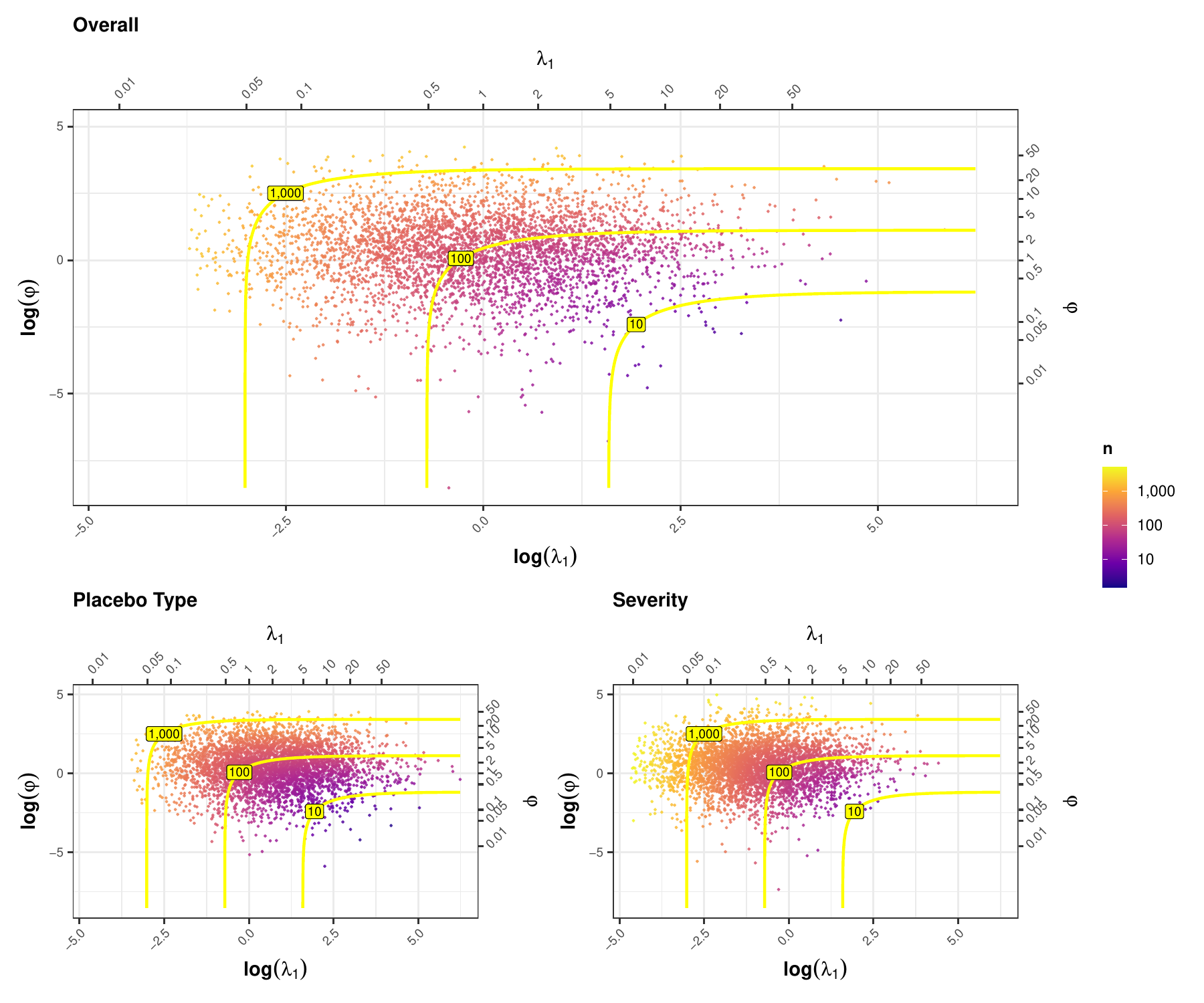}
    \caption{Posterior predictive joint distribution of the placebo exacerbation
    rate $\lambda_1(2017)$ and overdispersion parameter $\varphi$ for a new study
    in 2017 based on the covariate-based models. Results are shown for the overall
    dataset and for analyses incorporating placebo type or asthma severity as
    study-level covariates in the event-rate model. Each point represents a
    posterior predictive draw from the hierarchical model. Colours indicate the
    corresponding required sample size per group under the assumed rate ratio of
    0.5, two-sided $\alpha=0.05$, target power $1-\beta=0.80$, and follow-up
    duration $d=1$ year.}
    \label{fig:joint-sample-size-covariate}
\end{figure}

 \begin{figure}[H]
    \centering
    \includegraphics[width=1\linewidth]{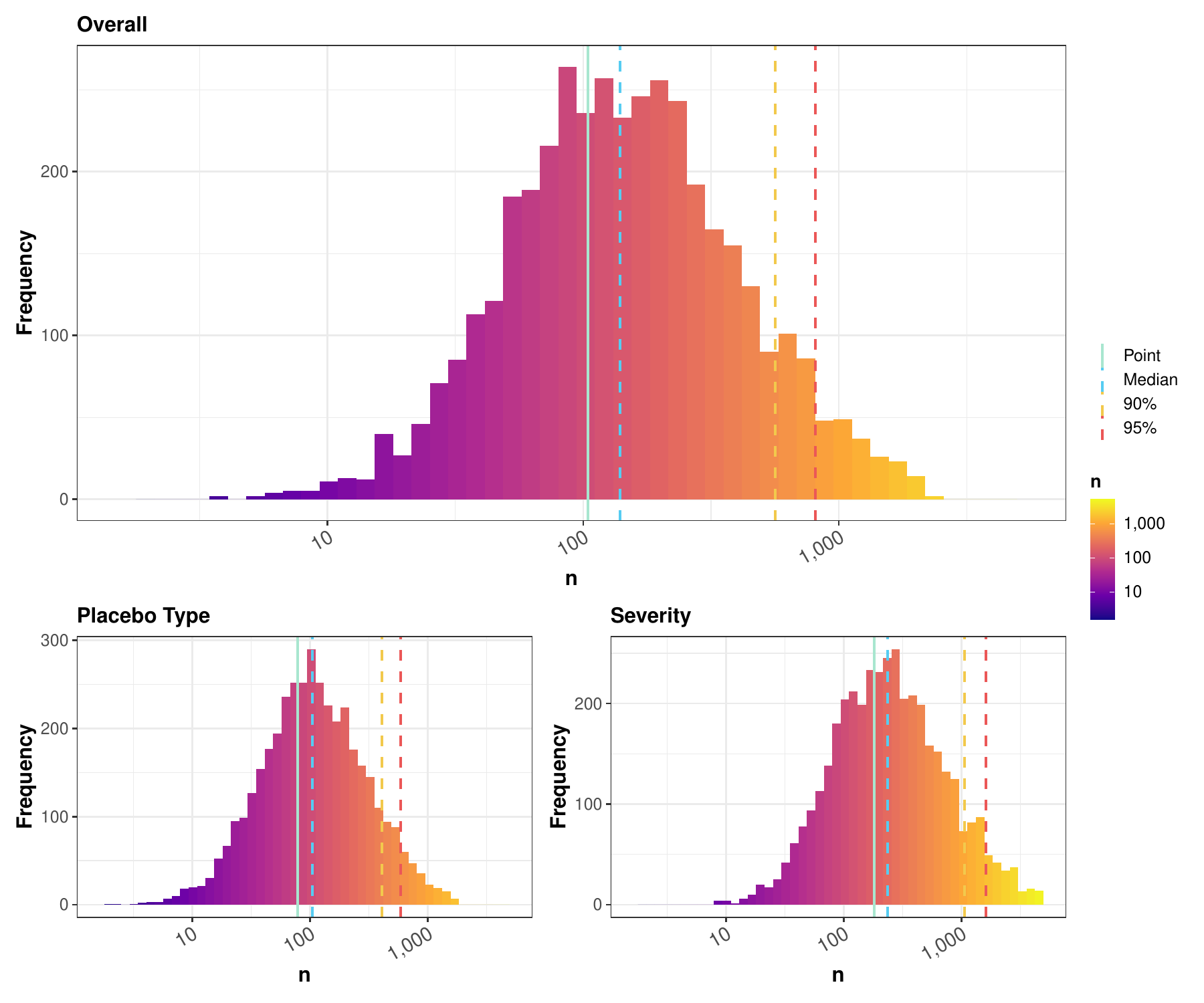}
    \caption{Posterior predictive distributions of the required sample size per
    group for a new study in 2017 based on the covariate-based analyses.
    Histograms show the distribution of sample sizes obtained by propagating
    posterior uncertainty in the baseline event rate $\lambda_1$ and overdispersion
    parameter $\varphi$, with placebo type or asthma severity incorporated as
    study-level covariates in the event-rate model. Vertical lines indicate the
    point estimate based on posterior medians, the posterior median sample size and
    the 90th percentile of the posterior predictive sample-size distribution. The
    design parameters are a rate ratio of 0.5, two-sided $\alpha=0.05$, target power
    $1-\beta=0.80$, and follow-up duration $d=1$ year.}
    \label{fig:hist-sample-size-covariate}
\end{figure}

\section{Study characteristics by placebo type and asthma severity}
\label{AppendixE}

\renewcommand{\thetable}{E\arabic{table}}
\setcounter{table}{0}
\renewcommand{\thefigure}{E\arabic{figure}}
\setcounter{figure}{0}

Table~\ref{table:baseline_subgroups} summarizes the characteristics of the
included studies separately by placebo type and asthma severity.
\begin{sidewaystable}[!htbp]
\centering
\scriptsize
\caption{Characteristics of the included studies according to placebo type and
asthma severity. Values are presented as medians and ranges across studies;
$N$ denotes the number of studies with available information.}
\label{table:baseline_subgroups}

\begin{tabular}{@{}lccc c ccc c ccc@{}}
\toprule

\multicolumn{12}{l}{\textit{Panel A: Placebo type}} \\[2pt]
& \multicolumn{3}{c}{True placebo} & &
  \multicolumn{3}{c}{ICS-placebo} & &
  \multicolumn{3}{c}{ICS--LABA-placebo} \\
\cmidrule(lr){2-4}
\cmidrule(lr){6-8}
\cmidrule(lr){10-12}
& $N$ & Median & Range & &
  $N$ & Median & Range & &
  $N$ & Median & Range \\
\midrule

No. of patients
& 27 & 138 & 48--692
& & 16 & 177 & 27--3568
& & 19 & 234 & 54--2087 \\

Study duration (yr)
& 27 & 0.500 & 0.067--1
& & 16 & 0.519 & 0.230--3
& & 19 & 0.923 & 0.250--1 \\

Males (\%)
& 27 & 40.0 & 27.1--61.65
& & 14 & 39.6 & 26.0--51.54
& & 18 & 37.9 & 25.9--49.0 \\

Mean age (yr)
& 27 & 43.6 & 26.0--53.9
& & 12 & 49.35 & 24.0--55.0
& & 16 & 48.15 & 36.0--54.0 \\

Mean \fevone{} (L)
& 17 & 2.210 & 1.630--3.389
& & 11 & 2.200 & 1.700--2.700
& & 12 & 1.835 & 1.558--2.782 \\

Mean \fevonepp{} (\%)
& 25 & 68.4 & 52.69--96.1
& & 15 & 68.0 & 55.2--96.4
& & 17 & 68.0 & 54.6--84.8 \\

Smokers (\%)
& 13 & 7.00 & 0.21--38
& & 6 & 13.5 & 0--34
& & 9 & 0.48 & 0--27.8 \\

Mean AQLQ
& 9 & 4.10 & 3.60--5.50
& & 7 & 4.30 & 4.00--5.03
& & 11 & 4.60 & 3.72--5.35 \\

Mean ec ($\mu$L$^{-1}$)
& 9 & 325.0 & 0.25--400
& & 6 & 355.0 & 200--688
& & 5 & 290.0 & 190--370 \\

\midrule
\multicolumn{12}{l}{\textit{Panel B: Asthma severity}} \\[2pt]
& \multicolumn{3}{c}{Mild asthma} & &
  \multicolumn{3}{c}{Moderate asthma} & &
  \multicolumn{3}{c}{Severe asthma} \\
\cmidrule(lr){2-4}
\cmidrule(lr){6-8}
\cmidrule(lr){10-12}
& $N$ & Median & Range & &
  $N$ & Median & Range & &
  $N$ & Median & Range \\
\midrule

No. of patients
& 16 & 202 & 54--3568
& & 32 & 156 & 27--925
& & 14 & 191 & 58--813 \\

Study duration (yr)
& 16 & 0.750 & 0.230--3
& & 32 & 0.534 & 0.067--1
& & 14 & 0.795 & 0.106--1 \\

Males (\%)
& 15 & 43.0 & 30.0--61.65
& & 31 & 38.0 & 25.9--52.0
& & 13 & 37.0 & 32.0--55.0 \\

Mean age (yr)
& 13 & 35.5 & 24.0--55.0
& & 29 & 44.0 & 31.3--54.0
& & 13 & 51.0 & 45.7--53.9 \\

Mean \fevone{} (L)
& 6 & 2.990 & 2.400--3.389
& & 22 & 2.205 & 1.690--2.782
& & 12 & 1.740 & 1.558--2.200 \\

Mean \fevonepp{} (\%)
& 15 & 84.6 & 67.7--96.4
& & 31 & 68.0 & 60.3--79.7
& & 11 & 57.8 & 52.69--59.9 \\

Smokers (\%)
& 8 & 2.00 & 0--16
& & 12 & 19.0 & 0--34
& & 8 & 3.215 & 0--38 \\

Mean AQLQ
& 4 & 5.235 & 4.30--5.27
& & 18 & 4.15 & 3.60--5.50
& & 5 & 4.10 & 3.90--4.70 \\

Mean ec ($\mu$L$^{-1}$)
& 0 & -- & --
& & 12 & 365.0 & 0.25--688
& & 8 & 302.5 & 227--380 \\

\bottomrule
\end{tabular}
\end{sidewaystable}
\end{appendices}

\newpage

\end{document}